\documentclass[amssymb, nofootinbib, superscriptaddress]{revtex4}
\usepackage{hyperref}
\usepackage{amsfonts,amssymb,amsmath,exscale,bbm}
\usepackage{footnpag} 
\usepackage[all,knot]{xy}  
\usepackage{color}
\usepackage{tikz} 
\usepackage{graphicx}
\usepackage{epsfig}

\usepackage{amsmath}
\usepackage{amsfonts}
\usepackage{graphicx}
\usepackage{psfrag}
\usepackage{array}
\usepackage{bbm}
\usepackage{hyperref}
\usepackage{amsthm}
\theoremstyle{definition}

\newcommand{\tr}{\mathrm{tr}}

\newcommand{\cA}{{\mathcal A}}

\newcommand{\cC}{{\mathcal C}}
\newcommand{\cD}{{\mathcal D}}

\newcommand{\cG}{{\mathcal G}}

\newcommand{\cL}{{\mathcal L}}

\newcommand{\cP}{{\mathcal P}}

\newcommand{\cT}{{\mathcal T}}

\newcommand{\g}{{\mathfrak{g}}}

\def\inv{{\mbox{\tiny -1}}}
\def\minus{{\mbox{\small -}}}
\def\plus{{\mbox{\tiny +}}}

\newcommand{\R}{\mathbb{R}}
\newcommand{\C}{\mathbb{C}}

\newcommand{\Z}{\mathbb{Z}}

\newcommand\beq{\begin{equation}}
\newcommand\eeq{\end{equation}}
\newcommand{\be}{\begin{equation}}
\newcommand{\ee}{\end{equation}}
\newcommand{\bes}{\begin{eqnarray}}
\newcommand{\ees}{\end{eqnarray}}
\newcommand{\mone}{^{\minus 1}}

\def\ov{\overline}

\def\eps{\epsilon}

\def\cc{{\cal C}}

\def\ot{{\,\otimes \,}}

\def\act{\rhd}

\def\vphi{{\varphi}}

\def\g{\mathfrak{g}}

\def\ie{{i.e. \/}}

\def\cf{{c.f. \/}}
\def\dr{{\rightarrow}}

\def\cop{{\bigtriangleup}}

  \def\cc{{\cal C}}    
     \def\mm{{\cal M}} \def\nn{{\nonumber}}

\def\tr{{\mathrm{tr}}}

\def\hphi{{\widehat \phi}}

\def\hpsi{{\widehat \psi}}

\def\act{{\, \triangleright\, }}

\newcommand{\su}{\mathfrak{su}}

\newcommand{\so}{\mathfrak{so}}
\newcommand{\SU}{\mathrm{SU}}

\newcommand{\U}{\mathrm{U}}

\newcommand{\SO}{\mathrm{SO}}

\newcommand{\ISO}{\mathrm{ISO}}

\def\extd{\mathrm {d}}
\newcommand{\e}{\epsilon}

\newcommand\acts\triangleright

\newcounter{letter} \newcounter{numeral} \newcounter{Numeral}

\newcommand\Tr{\mathrm{Tr}}
\def\vphi{\varphi}
\def\vphihat{\widehat{\varphi}}
\def\hvphi{\widehat{\varphi}}

\def\e{\mbox{e}}
\def\extd{\mathrm {d}}

\newcommand\maps{\colon}

\begin{document}

%

%

\title{\Large \bf Diffeomorphisms in  group field theories}

\author{Aristide Baratin}\email{abaratin@aei.mpg.de}
\affiliation{Triangle de la Physique, CPHT \'Ecole Polytechnique, IPhT Saclay, LPT Orsay, \\
 Laboratoire de Physique Th\'eorique, CNRS UMR 8627, Universit\'e Paris XI, F-91405 
Orsay C\'edex, France}


\author{Florian Girelli}\email{girelli@physics.usyd.edu.au}
\affiliation{School of Physics, University of Sydney, Sydney, New South Wales 2006, Australia}

\author{Daniele Oriti}\email{doriti@aei.mpg.de}
\affiliation{Max Planck Institute for Gravitational
Physics, Albert Einstein Institute, Am M\"uhlenberg 1,
14467 Golm, Germany}


\begin{abstract}
\bigskip
\noindent 
We study the issue of diffeomorphism symmetry in group field theories (GFT), using the noncommutative metric representation introduced in \cite{aristidedaniele}.  
In the colored Boulatov model for 3d gravity, we identify a field (quantum) symmetry  which ties together the vertex translation invariance of discrete gravity, 
the flatness constraint of canonical quantum gravity, and the topological (coarse-graining) identities for the 6j-symbols. 
We also show how, for the GFT graphs dual to manifolds,  the invariance of the Feynman amplitudes encodes the discrete residual action of diffeomorphisms in simplicial gravity path integrals. 
We extend the results to  GFT models for higher dimensional BF theories and discuss various insights that they provide on the GFT formalism itself. 
\end{abstract}

\maketitle
\bigskip \bigskip

%
\section{Introduction}
%

Diffeomorphism symmetry is a crucial aspect of the dynamics of spacetime geometry as described by general relativity 
and its higher derivative extensions. 
It is tied to the notion of  background independence \cite{back}, as the introduction of a non-dynamical background breaks the full diffeomorphism invariance. 
It also imposes strong constraints on the allowed dynamics. 
In fact,  for example, the only diffeomorphism invariant action (in 4d) for a tensor metric field that involves at most its first derivatives is the Einstein-Hilbert action (with cosmological constant); 
and, in a canonical formalism based on intrinsic metric and conjugate extrinsic curvature, only canonical general relativity is compatible with the algebra of (the canonical counterpart of) diffeomorphisms \cite{kucharteit}. 


This fact acquires even more relevance from the point of view of ongoing efforts to build a quantum theory of gravity. 
In background independent approaches \cite{libro}  aiming at explaining the very origin of spacetime geometry starting from `pre-geometric',  discrete or purely algebraic structures, the correct implementation of diffeomorphism invariance is a key guiding principle for the very definition of the microscopic dynamics. 
A major open problem in these approaches, such as in simplicial gravity \cite{renate}, spin foam models \cite{review}  and group field theories (GFT) \cite{gftreview},  
is to show how the dynamics reduce to general relativity in a semi-classical and continuum approximation. 
A good control over the (pre-geometric analogue of) diffeomorphism invariance is then essential: 
provided such an approximation does not break this symmetry, general relativity should emerge as the dynamics of the metric field defined in terms of the fundamental degrees of freedom of the theory, at least at leading order. 
If the invariance is only approximate, still the requirement that it becomes exact in the continuum limit is an important guiding principle for the definition of appropriate coarse-graining and renormalization procedures, or to identify the diffeomorphism invariant sector which should be dominant in the limit \cite{bianca}. 

With the smooth manifold of general relativity replaced by discrete structures, 
the issue becomes that of identifying suitable transformations of the pre-geometric data\footnote{In dynamical triangulations \cite{renate, DT}, all such data are fixed to constant values and the only analogue of diffeos is the automorphism group of the simplicial complex itself.} 
leaving the quantum amplitudes invariant and encoding the (residual) action of the diffeomorphism group.  
This is known in the context of Regge calculus \cite{regge}, 
where an action of diffeomorphisms at the vertices of the Regge triangulation has been shown to exist around flat solutions. 
This is understood geometrically as the invariance of the Regge action upon {\sl  translations} of the vertices, in a local flat embedding of the triangulation in $\R^d$.
The invariance is exact in 3d, where the geometry is constrained to be flat; it is only approximate in the 4d case, and in the presence of a cosmological constant (see \cite{bennybianca} and references therein). In both cases, the (approximate) invariance can be related to discrete 
Bianchi identities.
The action of diffeomorphisms in spin foam models has also been studied in the context of 3d gravity  \cite{laurentdiffeo}. 
In this work, it is shown that the discrete residual of the local Poincar\'e invariance, classically equivalent to diffeomorphism invariance, is responsible of (part of) the divergences of the Ponzano-Regge model. 
A related aspect of diffeomorphisms in spin foam models
is the algebraic expression of diffeomorphism invariance in terms of algebraic identities satisfied by $n$-$j$ symbols, at the root of the topological invariance of some models, and recognized to be an algebraic translation of the canonical gravity constraints \cite{BC-WdW, EteraValentin}.

\

Group field theories (GFT) \cite{gftreview}  are a higher dimensional generalization of matrix models \cite{mm}, and provide a second quantization of both spin network dynamics  and simplicial gravity. Their Feynman diagrams are dual to simplicial complexes,  the amplitudes are given equivalently as spin foam models or simplicial gravity path integrals \cite{aristidedaniele}.  Conversely, any spin foam model can be interpreted as a Feynman amplitude of a group field  theory \cite{carlomike}.  
Hence in the GFT perturbative expansion, one obtains a sum over (pre-)geometric data weighted by appropriate amplitudes, augmented by a sum over simplicial complexes of arbitrary topology. 
In this paper, we ask ourselves whether the various notions of diffeomorphisms invariance 
studied in the literature on discrete gravity can be traced back to a symmetry of the group field theory.  

This task had proven impossible to fulfill up to now. The main reason was the absence, at the GFT level, of explicit  metric variables, on which (discrete) diffeomorphisms would act. Now, recently, a metric formulation of GFT, completely equivalent to the usual formulations in terms of group variables or group representations, has been developed \cite{aristidedaniele} and used to prove an exact duality between spin foam models and simplicial path integrals. 
Here we use this formulation to study the action of discrete diffeomorphisms in GFT. By doing so, we relate in a clear way various aspects of diffeomorphism invariance in spin foam models, canonical loop quantum gravity and simplicial  gravity. 
More precisely, we show that there is a set of field transformations leaving the GFT action invariant, whose geometrical meaning in the various GFT representations ties together the symmetry of the Regge action and the simplicial Bianchi identities, the canonical constraints of loop quantum gravity (adapted to a simplicial complex) and algebraic identities satisfied by $n$-$j$ symbols. 

A key feature of this metric formulation, which recasts GFTs as  non-commutative field theories on Lie algebras, is to reveal and to make explicit the non-commutativity of the geometry in GFT and spin foam models \cite{SFnoncomm, etera, flux}. The action of discrete diffeomorphisms described in this paper naturally incorporates this non-commutativity, as it is generated by a Hopf algebra \cite{majid}.  
Diffeomorphism invariance in GFT thus takes the form of a deformed (quantum) symmetry. The definition of deformed symmetries in GFT, also considered in \cite{Floetera}, requires  to embed the field theory into the larger framework of braided quantum field theories \cite{oeckl}.

\

We work in the {\sl colored} version of the GFT formalism \cite{razvan,vincentcolored}, analogous to multi-matrix models \cite{mm}. 
The coloring can be used \cite{razvan2} to define a full homology for the GFT colored diagrams\footnote{For alternative definitions of homology of GFT diagrams, see \cite{ValentinMatteo}.} and to unambiguously associate to it a triangulated pseudo-manifold, that is, complexes with point-like topological singularities \cite{DP-Petronio}.
The color formalism eliminates more pathological diagrams that are instead generated by standard GFT's \cite{razvan}. 
Strikingly, the coloring turns out to be also crucial for recasting the perturbative expansion of the (colored) Boulatov model, with a cut-off in representation space, in terms of  a topological expansion, and to show that the sum is dominated by manifolds of trivial topology in the large cut-off limit \cite{RazvanN}. 
This is the GFT analogue of the `large-N'  expansion of matrix models. These are very strong motivations for introducing coloring in  GFT models. 
In this paper we give another one: it is {\it only} in the colored framework that the action of discrete diffeomorphisms can be encoded into field transformations. 

We focus on the topological models, namely the (colored) Boulatov and Ooguri models  for 3d gravity and 4d BF theory. 
The analysis can however be extended to 4d gravity models obtained by imposing constraints on topological ones \cite{4ddiffeo}. 

\

The paper is organized as follows. 
In Section \ref{introGFT}, we review the GFT framework in dimension 3, in its three known formulations: 
the `group' formulation in terms of fields on a group manifold, the `spin' formulation in terms of tensors in group representations, and the recent `metric' formulation  in terms of fields on Lie algebras. We illustrate how the duality of GFT representations translates into an exact duality between spin foam models, lattice gauge theory 
and simplicial path integrals. 

In Section \ref{Fieldsymmetry}, we introduce a set of field transformations which, we show, leave invariant  the action of the colored Boulatov for 3d gravity. 
These transformations are generated by a Hopf algebra \cite{majid}, more precisely by the translational part of a deformation of the Poincar\'e group. 
The definition of deformed (quantum) symmetries on GFT requires to embed the field theory into the larger framework of braided quantum field theories \cite{oeckl}. 
We exploit the invariance of the GFT vertex function to give the  geometrical meaning of the symmetry  in the three GFT representations. We find that: 
\begin{enumerate}
\item in the `metric' representation, the symmetry reflects the invariance under translations of each of the vertices of the Euclidean tetrahedron patterned the GFT interaction.\vspace{-0.2cm}
\item in the `group' representation, the symmetry expresses the flatness of the boundary connection that  the field variables represent.  
\vspace{-0.2cm}
\item in the `spin' representation, the symmetry encodes the topological identities and recursions relations of the 6j-symbols. 
\end{enumerate} 

In Section \ref{GFTvsSFsymmetries}, we look at the invariance of the GFT amplitudes and explain how the GFT symmetry relates to the action of diffeomorphisms in simplicial path integrals. The analysis naturally distinguishes between manifold graphs and pseudo-manifold ones.  
In the case of manifold graphs, we show, both geometrically and algebraically, how to derive discrete Bianchi identities from the invariance of the vertex and propagators functions. 

Finally, in Section \ref{4dBFcase}, we extend the results to the GFT model for 4d BF theory and discuss the case of constrained models for gravity. 
We conclude in Section \ref{additional} with a discussion of various issues raised by our analysis and new insights that it provides on the GFT formalism. 

\section{Colored GFT's and metric representation} 
\label{introGFT} 
d-Dimensional GFT's \cite{gftreview},  in their {\sl colored} version \cite{razvan}  are field theories described in terms of d+1 complex  fields $\{\vphi_\ell\}_{\ell=1\cdots d+1}$ defined over d copies of a group $G$, with a certain gauge invariance. The index $\ell$ is referred to as the color of the fields.
Here we consider the 3d case and the Euclidean rotation group 
$G=\SO(3)$, so that each field $\varphi_\ell$ is  a function on $\SO(3)^{\ot 3}$. 
The gauge invariance condition reads:
\beq \label{gauge}
\forall h \in \SO(3),  \qquad \varphi_\ell(hg_1, hg_2, hg_3)  \, = \, \varphi_\ell(g_1, g_2, g_3) .
\eeq
\begin{figure}
\begin{center}
\includegraphics[scale=.4]{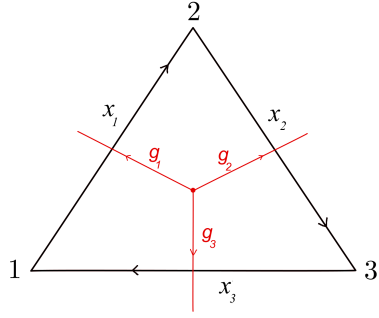}
\label{triangle}
\caption{Geometric interpretation of the GFT field.}
\end{center}
\end{figure}
The dynamics is governed by the action $S[\vphi]= S_{kin} [\vphi] + S_{int}[\vphi]$, where the kinetic term couples fields with the same colors: 
\beq \label{kinetic}
S_{kin} [\vphi] = \int [\extd g_i]^3 \sum_{\ell=1}^4   \, \vphi_\ell(g_1, g_2, g_3) \overline{\vphi_{\ell}}(g_1, g_2. g_3) ,
\eeq
where $[\extd g]^n$ is the product Haar measure on the group $\SO(3)^{\ot n}$, and $\overline{\vphi_{\ell}}$ are the complex conjugated  fields. The interaction is homogeneous of degree 4 and given by 
\bes \label{interaction1}
S_{int}[\vphi] &=& \lambda \int [\extd g_{i} ]^6 \, \vphi_1(g_1, g_2, g_3) \vphi_2(g_3, g_4, g_5) \vphi_3(g_5, g_2, g_6) \vphi_4(g_6, g_4, g_1)
\nn \\
& & +\lambda \int [\extd g_{i} ]^6 \, \overline{\vphi_4}(g_1, g_4, g_6) \overline{\vphi_3}(g_6, g_2, g_5) \overline{\vphi_2}(g_5, g_4, g_3) \overline{\vphi_1}(g_3, g_2, g_1). 
\ees
The six integration variables in each integral follows the pattern of the edges of a tetrahedron. A field represents a triangle, 
the three field arguments being associated to its edges (see Fig. 1).  
The four triangles  of the tetrahedron are marked by distinct colors\footnote{The coloring of each field, and thus of each triangle, by a single label $\ell$ can be equivalently converted in a coloring of each vertex of the tetrahedron by a label in the same range. In this setting, each field-triangle is labelled by the three colors of its three vertices. This shows that colored GFTs are a field theory generalization of double-indexed 3d tensor models \cite{3dtensors}.}. 
When the fields with different colors are all identified $\vphi_{\ell} \!:=\! \vphi$ to a single real field,  colored GFT's reduce to standard GFT's. 

The Feynman expansion of a GFT generates stranded diagrams, with 3 strands per propagator, equipped with a canonical orientation of all lines and higher dimensional faces.  The propagator and vertex for $\vphi$ are drawn in Fig. 2; the vertex for $\overline{\vphi}$ is obtained by reversing the order of all labels.  
While the interaction vertex patterns a tetrahedron with colored triangles, the propagator glues together tetrahedra  along triangles of the same color. 

\begin{figure}
\begin{center}
\includegraphics[scale=.4]{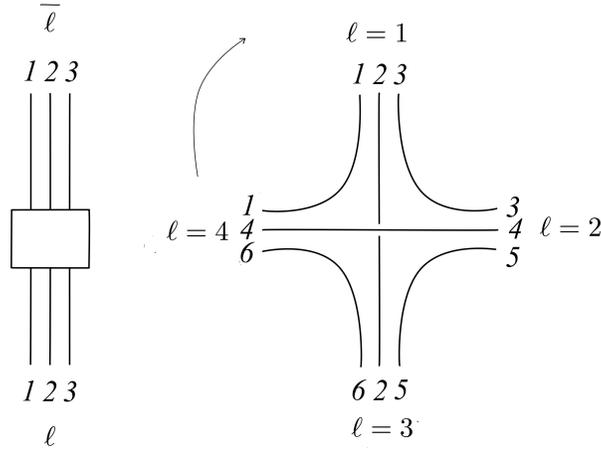} 
\label{Feynmanrules}
\caption{3D GFT propagator and vertex. } 
\end{center}
\end{figure}

\begin{figure}
\begin{center}
\includegraphics[scale=.4]{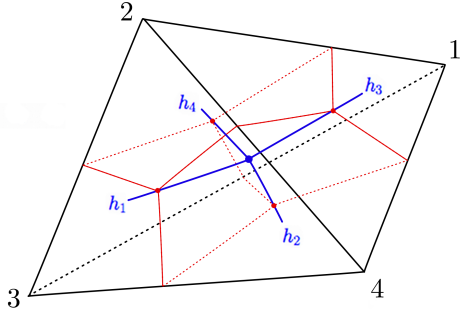} 
\caption{ The `boundary' holonomies $g^\ell_i$ are in red, while the `bulk' holonomies $h_\ell$ are in blue.}  \label{interaction}
\end{center}
\end{figure}

Graph amplitudes are build out of propagators and vertex functions: 
\beq \label{Feynmanfunctions}
P_\ell(g, g') = \int \extd h \prod_{i=1}^3 \delta(g^{-1}_i h g'_i),  \qquad 
V(g, g') = \int \prod_{\ell=1}^4 \extd h_{\ell} \prod_{i=1}^6 \delta((g^{\ell}_i)^{-1} h_\ell h^{-1}_{\ell'} g^{\ell'}_i),
\eeq 
which identifies the variables along connected strands, modulo left shift by the gauge variables $h$ arising from the invariance (\ref{gauge}). 
The  vertex function has an interpretation in terms of lattice gauge theory, where the three group variables $g^\ell_i$ and the group variables $h_\ell$ are viewed as holonomies along the links of the complex topologically dual to a tetrahedron, shown in Fig. \ref{interaction}.  
The $g^\ell_i$ are `boundary' holonomies along the links dual to a triangle $\ell$. The $h_\ell$ are `bulk' holonomies along the links connecting the triangles to the center of the tetrahedron. The vertex function simply states that the two dimensional faces of the complex dual (in red in Fig. \ref{interaction}) are flat. This implies that the encoding of geometric information in the model fits a piecewise-flat context, as in simplicial quantum gravity approaches.

In gluing two tetrahedra, the propagator function identifies the boundary variables of the shared triangle, up to a group variable $h$ interpreted as a  further parallel transport through  the triangle.

After integration over all boundary variables $g$, the amplitude of  a closed GFT diagram $\cG$ takes the form
\beq \label{PRlattice}
\cA_\cG= \int \prod_l \extd h_l \, \prod_f \, \delta \left( {\overrightarrow{\prod}}_{l \in \partial f} \, h_l \right)
\eeq
where the products are over the lines $l$ and the faces  (loops of strands) $f$ of the diagram.  $l$ and $f$ dually label the triangles and the edges of the triangulation defined by the diagram. The ordered products of line variables along the boundary $\partial  f$ of the faces are computed by choosing an orientation and a reference vertex for each face $f$. The group variables are taken to be $h_l$ or $h_l^{-1}$, depending on whether the orientations of $l$ and $f$ agree or not. In terms of lattice gauge theory, the set of variables $(h_l)_{l\in \cG}$ gives a discrete connection on (the complex dual to) the triangulation, giving parallel transports from a tetrahedron to another.  The delta functions in (\ref{PRlattice}) imposes this connection to be flat. The model is already seen then as describing a discrete version of topological 3d BF theory, discretized on the simplicial complex dual to the GFT diagram.

The `spin' representation of the GFT is obtained using the Peter-Weyl expansion of the fields over half-integer spins labeling the representations of $\SO(3)^{\ot 3}$. 
Because of gauge invariance, the coefficients are proportional to the $\SO(3)$ Clebsh-Gordan coefficients $C^{j_1, j_2, j_3}_{m_1, m_2,m_3}$; 
the interaction vertex is  expressed in terms of  6j-symbols.  The amplitude of a Feynman diagram gives the Ponzano-Regge spin foam model \cite{PonzanoRegge}:
\beq \label{PRspin}
\cA_\cG= \sum_{\{j_e\}}  \prod_e d_{j_e} \prod_{\tau} \left\{\begin{array}{ccc}
j_1^\tau&j_2^\tau&j_3^\tau \\
j_{4}^\tau&j_{5}^\tau&j_{6}^\tau
\end{array}
\right\},
\eeq
where the spins $j_e$ label the edges of the triangulation associated to the diagram, $d_j = 2j+1$ is the dimension of the representation $j$, and the amplitude is a product of tetrahedral 6j-symbols. Thus,  group and spin representations of the GFT realize explicitly the duality between the connection  (\ref{PRlattice}) and spin foam  (\ref{PRspin}) formulations of the Ponzano-Regge model \cite{review, PR1}.

A third representation of GFT's in terms of continuous non-commutative `metric' variables $x \in \su(2) \sim \R^3$, has been recently developed \cite{aristidedaniele} and shown to realize a further duality between spin foam models and simplicial path integrals. Since the geometrical meaning of the symmetries studied in the next sections is best understood in such a metric representation,  let us briefly recall here its construction. 
The representation is obtained using the group Fourier transform \cite{majid, SFnoncomm} of the  fields
\beq \label{fourier}
\vphihat_\ell(x_1, x_2, x_3) :=\int  [\extd g_i]^3\, \vphi_\ell(g_1, g_2, g_3)  \,\e_{g_1}(x_1) \e_{g_2}(x_2) \e_{g_3}(x_3) ,
\eeq
expressed in terms of plane-wave functions $\e_g \maps \su(2) \!\sim\! \R^3 \to \U(1)$. The definition of the plane-wave depends on a choice of coordinates systems on the group manifold. In the following we  identify functions of $\SO(3) \!\sim\!\SU(2)/\Z_2$ with functions of $\SU(2)$ invariant under $g \to - g$. 
Using the parametrization $g \!=\! e^{\theta \vec{n}\cdot \vec{\tau}}$ and $x \!=\! \vec{x} \cdot \vec{\tau}$ of group and Lie algebra elements in terms of the (anti-Hermitian) $\su(2)$ generators $\vec{\tau} \!=(\! \tau_1, \tau_2, \tau_3)$,  a convenient representation of the plane-waves is 
\be
\e_{g}(x) := e^{i \Tr x g},
\ee
where the trace is given by $\Tr \tau_i \tau_j = - \delta_{ij}$. Note that, since $\overline{\e_{g}}(x) = \e_{g^\inv}(x) = \e_g(-x)$, the  Fourier transform 
of the complex conjugate field relates to the complex conjugate of the Fourier transform as 
\beq \widehat{\overline{\vphi}}_{\ell}(x_1, x_2, x_3) = \overline{\vphihat}_{\ell}(-x_1, -x_2, -x_3).\eeq 

The image of the Fourier transform inherits by duality a non-trivial (non-commutative) pointwise product from the convolution product on the group.  It is defined on plane-waves as 
\beq  \label{star}
(\e_{g} \star \e_{g'})(x) \!:=\! \e_{gg'}(x),
\eeq 
extends component-wise to the product of three plane-waves  and by linearity to the whole image of the Fourier transform.  

The first feature of this representation is that the gauge invariance condition (\ref{gauge}) expresses itself as a `closure constraint' for the triple of variables $x_i$ of the dual field. To see this, we consider the projector $\cP$ onto gauge invariant fields:
\beq\label{projector}
\cP \act \vphi_\ell = \int [\extd h]\, \vphi_\ell(h g_1, h g_2, h g_3),
\eeq
and note that
\beq
\widehat{\cP\act\vphi_\ell} = \widehat{C} \star \vphihat_\ell, \qquad \widehat{C}(x_1, x_2, x_3) := \delta_0(x_1 \!+\!x_2\!+\!x_3) ,
\eeq
where $\delta_0$ is the element $x=0$ of the family of functions
\beq \label{delta}
\delta_x(y)  := \int [\extd h] \, \e_{h^\inv}(x)  \e_h(y).
\eeq
These functions play the role of Dirac distributions in the non-commutative setting, as
\beq \label{Dirac}
\int [\extd^3 y] \, (\delta_x \star f)(y)= f(x) ,
\eeq
where $\extd^3 y$ is the standard Lebesgue measure on $\R^3$. 
We may thus interpret the variables $x_i$ of the  dual gauge invariant  field as the closed edges vectors of a triangle in $\R^3$, and further confirm the interpretation of the GFT fields as (non-commutative) triangles. 

The GFT action can be written in terms of dual fields and metric variables by exploiting the duality between group convolution and $\star$-product. 
Given two functions $f, h$ on $\SO(3)$ and $\widehat{f}, \widehat{h}$ their Fourier transform \eqref{fourier}, this duality can be read in the 
property 
\beq \label{dualproduct}
\int [\extd g]\, f(g) h(g) = \int [\extd^3 x ]\, (\widehat{f} \star \widehat{h}_{\minus})(x),
\eeq
where $\widehat{h}_\minus(x) := \widehat{h}(-x)$ and $\extd^3 x$ is the Lebesgue measure on $\R^3$.
Hence the combinatorial structure of the GFT action in the metric representation is the same as in group one, while group convolution is replaced by $\star$-product.    Using the short notation $\vphihat_\ell^{123}:= \vphihat_\ell(x_1, x_2, x_3)$, we can write the action as 
\bes \label{action in x}
S [\vphihat]  &=& \int [\extd^3 x_i]^3 \,\sum_{\ell=1}^4   \, \vphihat_\ell^{123} \star \widehat{\overline{\vphi}}_\ell^{\lower0.1in \hbox{\footnotesize123} } \nn \\
&+& \lambda  \int  [\extd^3 x_i]^6\, \vphihat_1^{123} \star \vphihat_2^{345}\star \vphihat_3^{526} \star \vphihat_4^{641}
+\lambda\int  [\extd^3 x_i]^6 \, \widehat{\overline{\vphi}}_4^{146} \star \widehat{\overline{\vphi}}_3^{345}\star \widehat{\overline{\vphi}}_2^{526}\star  \widehat{\overline{\vphi}}_1^{641},
\ees
where it is understood that $\star$-products relate repeated upper indices as $\hvphi^i \star \hvphi^i \! :=\! (\hvphi \star \hvphi_{\minus})(x_i)$, with $\hvphi_{\minus}(x) = \hvphi(- x)$.  

Feynman amplitudes are build out of propagators and vertex functions: 
\beq \label{Feynmanfunctions2}
P_\ell(x, x') = \int [\extd h]\, \prod_{i=1}^3 (\delta_{\minus x_i} \star \, \e_{h} )(x'_i),   \qquad 
V(x, x') = \int \prod_{\ell=1}^4 [\extd h_{\ell}]\, \prod_{i=1}^6 ( \delta_{\minus x^{\ell}_i} \star\, \e_{h_\ell h^{-1}_{\ell'}})(x^{\ell'}_i)
\eeq
where the $\delta_x$ are given by (\ref{delta}). These have a natural interpretation in terms of simplicial geometry, where the $x$ variables on connected strands
encode the metric of the same edge in different frames,  related with each other by the holonomies $h$.
In building up the diagram, propagator and vertex strands are joined to one another using the $\star$-product. 

Under the integration over the holonomy variables,  the amplitude of a closed diagram $\cG$ factorizes into a product of face amplitudes $A_{f}[h]$ taking the form of a cyclic $\star$-product:  
\beq \label{Feyn-integrand}
\cA_{f}[h] = \int \prod_{j=0}^{N_f} [\extd^3 x_j]\, \vec{\bigstar}_{j=0}^{N + 1} \, (\delta_{x_j} \star \e_{h_{jj+1}})(x_{j+1}) 
\eeq
where the product is over the $N_f$ vertices of $\cG$ (dual to tetrahedra) in the loop of strands that bound $f$.  The ordered $\star$-product is computed by choosing an orientation and a reference vertex for the face $f$; by convention we set  $x_{N+1} :=x_0$. 
The holonomy $h_{jj\plus1}$
parallel transports the reference frame of $j$ to that of $j+1$. 
In terms of simplicial geometry, it encodes the identification, up to parallel transport, of the metric variables associated to the edge dual to $f$ in the different frames $j$.
  
After integration, within  all face amplitudes, over all metric variables $x_j$ except for that $x_0$ of the reference frame, the amplitude of the GFT diagram $\cG$ takes the form of a simplicial path integral: 
\beq \label{PRbf} 
\cA_\cG = \int \prod_l [\extd h_l]\, \prod_e [\extd^3 x_e]  \, e^{i \sum_e \Tr \, x_e H_e},
\eeq
where the products are over the lines of $\cG$ and the edges of the dual triangulation, and $H_e :={\overrightarrow{\prod}}_{l \in \partial f_e} \, h_l$ is the holonomy along the boundary of the face $f_e$ of $\cG$ dual to $e$, calculated from a given reference tetrahedron frame. 
The exponential term is the (exponential of the) discrete action of first order 3D gravity (which is the same as 3D BF theory), in Euclidean signature.  This gives the definite confirmation of the interpretation of the $x_e$ variables as discrete triad variables associated to the edges of the triangulation dual to the GFT Feynman diagram (edge vectors).

Thus, the metric representation of GFT realizes explicitly the duality between spin foam models (\ref{PRlattice}) or (\ref{PRspin}) and simplicial path integrals (\ref{PRbf}), generalizing it to arbitrary transition amplitudes (corresponding to open GFT diagrams) with appropriate boundary terms arising naturally in the simplicial action in (\ref{PRbf}), for fixed triad variables at the boundary, and boundary  observables.  This result is general: it extends to BF theories in higher dimension,  and to gravity models obtained as constrained BF theories (see \cite{aristidedaniele, aristidedanieleBC}). 

The metric representation has of course the advantage of making the (non-commutative, simplicial) geometry of GFT and spin foam models more transparent. 
This will be useful for the understanding of the symmetries studied in the next section.

\section{GFT (discrete) diffeomorphism symmetry} 
\label{Fieldsymmetry}

In this section, we introduce a set of field transformations which,  we show, leave the GFT action invariant. 
We give the geometrical meaning of such transformations in the different representations, and show that they correspond to diffeomorphisms in discrete quantum geometry models. We also derive yet another GFT representation in terms of the generators of the symmetry, 
which are Lie algebra `position' variables associated to the {\it vertices} of the simplex patterned by the GFT field.

The non-commutativity of the metric (triad) space plays a crucial role in the definition and meaning of the symmetry transformation. This is in fact 
a Hopf algebra (quantum) symmetry, characterized by a non-trivial action on a tensor product of fields, due to a non-trivial coproduct. The relevant quantum group here, i.e. in this specific GFT model for Euclidean 3d gravity with local gauge group being $\SO(3)$, is a deformation of  the  Euclidean group $\ISO(3)$, the so-called Drinfeld double $\cD \SO(3)$\footnote{The role played by the Drinfeld double in spin foam and GFT models has been emphasized already, e.g. in 
\cite{PRII, karimDSU, Floetera}.} .


\subsection{Action of  $\cD \SO(3)$ on fields on $\SO(3)$} 
\label{sec:dsu2}

The Drinfeld double is defined as $\cD \SO(3)\!=\!\cc(\SO(3))\rtimes \C\SO(3) $, where the group algebra $\C\SO(3) $ acts by the adjoint action on  the algebra of functions $\cc(\SO(3))$. It is a deformation of the three dimensional Euclidean group $\ISO(3)$ -- more precisely of the Hopf algebra $\cc(\R^3)\rtimes \C\SO(3) $ --  where the `rotations'  belong to the group algebra $\C\SO(3)$ and  the `translations'  are complex functions in    $\cc(\SO(3))$. A general element can be written as a linear combination of elements $f \otimes \Lambda$, where $f\in\cc(\SO(3))$ and $\Lambda \in \SO(3)$. 
The space of functions on the group $\cc(\SO(3))$ gives a representation of $\cD \SO(3)$,  in which rotations act by adjoint action on the variable and translations act by multiplication: 
\be \label{rep}
\phi(g)\,\mapsto \, \phi(\Lambda\mone \act g) \!:=\!\phi(\Lambda\mone g \Lambda), \qquad \phi(g)\,\mapsto\, f(g)\phi(g), \quad \phi\in \cc(\SO(3)).
\ee 
Choosing as translation element a generating plane-wave labelled by  $\varepsilon \in \su(2) \sim \R^3$, 
the field $\phi$ gets multiplied by a phase  $f_\varepsilon(g)=\e_g(\varepsilon)$. Upon the group Fourier transform introduced in the previous section,
\be
\hphi(x) = \int [\extd g]\, \phi(g) \,\e_g(x),\quad 
\ee
this corresponds to the dual action $ \hphi(x) \mapsto \hphi(x + \varepsilon)$. We will also use the {\sl dual} action of $\cD \SO(3)$, 
where rotations act by inverse adjoint action and plane-waves labelled by $\varepsilon$ acts by translation by $-\varepsilon$.

Up to now, the transformations are the exact analogue of the usual Poincar\'e transformations on functions on flat space, here replaced by the algebra $\su(2)$ while momentum space is replaced by the group manifold $\SO(3)$.  The deformation manifests itself as a non-trivial action on a tensor product of fields, due to the non-trivial co-product  on the translation algebra $\cc(\SO(3))$. Thus:
\be
\phi_1(g_1) \phi_2(g_2) \, \mapsto \, \cop f(g_1 \ot g_2) \phi_1(g_1)\phi_2(g_2), 
\label{act}
\ee
 where the coproduct $\cop$ is given by 
\be\label{cop CG}
\cop f (g_1\ot g_2)=f_{(1)}(g_1) f_{(2)}(g_2)= f(g_1g_2), \quad \forall f\in \cc(\SO(3)).
\ee
Using the fact that $\e_{g_1g_2}(\varepsilon) \!=\! (\e_{g_1} \star \e_{g_2})(\varepsilon)$, one can check that the dual action 
of the plane-wave $\e_g(\varepsilon)$ on a tensor product is obtained by translating each  variables by $\varepsilon$ and by taking the $\star$-product of the resulting fields with respect to $\varepsilon$:
\beq
\hphi_1(x_1) \hphi_2(x_2) \, \mapsto \, \hphi_1(x_1+ \varepsilon) \star_\varepsilon \hphi_2(x_2+ \varepsilon) .
\eeq
This structure is what replaces the usual translation group $\R^3$, and the deformation is consistent with the non-commutativity of the algebra of functions on $\su(2) \sim \R^3$ induced by  the $\star$-product. 

\subsection{GFT as a braided quantum field theory}

In order to allow the Hopf algebra to act on the polynomials of fields defined by the GFT action, the idea is to embed the theory into the algebraic framework of braided quantum field theories defined by Oeckl \cite{oeckl}.  
In short, this consists of lifting all polynomials of fields 
to tensor products, 
in order to keep  track of the ordering of the fields and field variables. 
Commuting  fields or field variables requires to specify  {\sl braiding} maps $B_{12} \maps X_1 \otimes X_2 \to X_2 \otimes X_1$ 
between any two copies of the space of fields. 
The theory is defined perturbatively as a {\sl braided} Feynman diagram expansion, using   a {\sl braided} Wick theorem \cite{oeckl}. 
In a trivial embedding of GFT's, where all fields commute, into the braided framework, the braiding maps are chosen to be the trivial  flip maps:
\bes 
B_{12}:&& \cc(\SO(3))\ot \cc(\SO(3)) \to \cc(\SO(3))\ot \cc(\SO(3))\nn\\
&&\phi(g_1)\ot \phi(g_2) \mapsto  \phi(g_2)\ot \phi(g_1) \nn
\ees
We emphasize that such trivial embedding does {\sl not} modify the theory\footnote{In fact, in this setting, the braided Feynmanology is redundant and the braided amplitudes coincide with the unbraided ones.}. It however allows us define Hopf algebra transformations on the GFT fields.  

The possibility of using a non-trivial braiding between fields or field arguments is not employed in usual GFTs, 
so we will stick to the usual formalism in what follows. The choice of trivial braiding is often made also in the non-commutative geometry literature, even in the presence of quantum group symmetries. However, since in general the trivial braiding  map does not intertwine the action of the quantum group symmetry, 
this choice leads to a breaking of the symmetry at the level of the n-point functions. In order to make the full theory symmetric, it is most natural to use the braiding of the (braided) category of representations of the quantum group \cite{majid}. Thus, in scalar field theory fully invariant under the Drinfeld double $\cD\SO(3)$, this braiding is
\bes
B_{12}:& \cc(\SO(3))\ot \cc(\SO(3))\dr \cc(\SO(3))\ot \cc(\SO(3))\nn\\
&\phi(g_1)\ot \phi(g_2)\mapsto  \phi(g_2)\ot \phi(g_2g_1g_2\mone) \label{braiding}
\ees
Moreover, the use of a trivial braiding in the presence of quantum group symmetries has been argued to be the origin of the (in-)famous IR-UV mixing \cite{sasai, oeckl2, mangano}.  Thus, the choice of  braiding does affect the physics of the model and modifies  its Feynman amplitudes.
 
The view we take in this paper opens the way to  a generalization of GFT's which would include a non-trivial braiding  intertwining the quantum symmetry described below. We discuss this  possibility in the concluding section. The suggestion of extending GFT's to fully braided field theories has been put forward also in the recent work \cite{Floetera},  again following the identification of a quantum group symmetry in the GFT context.

\subsection{Symmetries of the GFT action} 
\label{sec:sym}

We have recalled above how $\cD\SO(3)$ acts on a function of a single variable. Here we define a set of transformations of the GFT fields $\varphi_\ell$ under rotations and translations which leave the GFT action invariant. As we illustrate in the different GFT representations in the next subsection, the {\sl translational} part of this action, interpreted as `vertex translations' in the   simplex patterned by the field, will encode the action of discrete diffeomorphisms in GFT. 

Let us first point out that the requirement of gauge covariance restricts the number of independent transformations 
that a field transformation $\cT$ can undergo. 
Such transformation is indeed well-defined on gauge invariant field only if it commutes with the projector \eqref{projector}:
\[ 
 \cP\act (\cT\act \vphi_\ell) = \cT\act (\cP\act \vphi_\ell) .
\]
Thus, for instance, the only gauge covariant action of the rotations in $\cC\SO(3)^{\otimes 3}$:
\beq 
\vphi_\ell (g_1,g_2,g_3) \, \mapsto \, \vphi_\ell (\Lambda_{1,\ell} \mone \act g_1,\, \Lambda_{2,\ell}\mone \act g_2\,\Lambda_{3,\ell}\mone \act g_3), \label{rot}
\eeq
is the diagonal one $\Lambda_{i, \ell}:= \Lambda_{\ell}$.
In the metric representation, gauge covariance simply means that the transformation preserves  the closure $\delta(x_1^\ell + x_2^\ell + x_3^\ell)$ 
of the triangle $\ell$. In the case of rotations, one can easily go one step further and check that the only field transformation that preserves the kinetic and interaction polynomials is generated by a single rotation $\Lambda_\ell \!:=\!\Lambda$. 
In the metric representation, this is the only action of the rotations that respects the gluing of edge-vectors of the tetrahedron patterned by the interaction.

Let us stress that this symmetry corresponds precisely to the invariance under local changes of frame in each tetrahedron and in each triangle that one expects in 3d simplicial gravity (see Sec. \ref{GFTvsSFsymmetries}).  We thus find such an invariance implemented as the well-known local gauge invariance in both the simplicial path integral and pure gauge theory formulation of the GFT Feynman amplitudes, as well as in their spin foam representation.

\begin{figure}[h]
\begin{center}
\includegraphics[scale=.2]{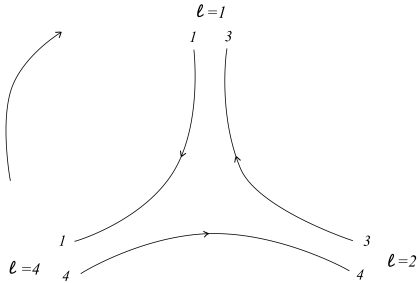} 
\caption{Vertex graph for the vertex $v_3$  }\label{3-vertex}
\end{center}
\end{figure}

We now turn to the more interesting case of translations.  We will define transformations generated by four $\su(2)$-translation parameters $\varepsilon_v$, where $v$ labels the four vertices of the interaction tetrahedron, diagrammatically represented by its dual diagram in Fig. 2.  
Each vertex  of this tetrahedron is represented by a certain subgraph, which we call `vertex  graph' \cite{FGO, razvan2}: the vertex graph for the vertex $v_\ell$ opposite to the triangle of color $\ell$ is obtained by removing all the lines which contain strands of color $\ell$.  The vertex graph of $v_3$ is pictured Fig. \ref{3-vertex}: 
its three lines pattern the three edges $1, 3, 4$ sharing  $v_3$. 

\noindent The vertices opposite to the triangles $\ell = 2, 3, 4$ are represented by identical (after anti-clockwise rotation by $\pi/4$, $\pi/2$ and $3\pi/4$) diagrams, where 1, 3, 4 are replaced respectively by $3, 5, 1$, by $5, 6, 4$ and by $6, 1, 2$  (in this order).
 
To define the action of a translation of the vertex $v_3$, we equip the lines of the vertex graph with an orientation, as drawn in the figure. Using this convention, each line has an `incoming' and an `outcoming' external strand. 
A translation of $v_3$ generated by  $\varepsilon_3 \in \su(2)$ acts non-trivially only on the strands of the vertex graph. 
In the metric representation, it  shifts the corresponding  variables $x_i^\ell$ by $\pm \varepsilon_3$ whether the strand $i$ comes in or out of $\ell$:
\bes \label{VertexTranslation0}
x_i^\ell &\mapsto& x_i^\ell + \varepsilon_3 \quad \mbox{if $i$ outgoing} , \\
x_i^\ell &\mapsto& x_i^\ell - \varepsilon_3 \quad  \mbox{if $i$ incoming}\nn.
\ees
in a way that preserves the closure $\delta(x_1^\ell + x_2^\ell + x_3^\ell)$ of each triangle $\ell$.  
More precisely, the translation $\cT_{\epsilon_3}$ of the vertex $v_3$ acts on the dual fields as:
\bes \label{VertexTranslation1}
\cT_{\varepsilon_3} \act \widehat\varphi_1(x_1, x_2, x_3) &:=& \bigstar_{\varepsilon_3} \, \widehat\varphi_1(x_1-\varepsilon_3, x_2, x_3+\varepsilon_3)   \nn \\
\cT_{\varepsilon_3} \act \widehat\varphi_2(x_3, x_4, x_5) &:=& \bigstar_{\varepsilon_3} \, \widehat\varphi_2(x_3-\varepsilon_3, x_4+\varepsilon_3, x_5) \nn \\
\cT_{\varepsilon_3} \act \widehat\varphi_4(x_6, x_4, x_1) &:=&\bigstar_{\varepsilon_3} \, \widehat\varphi_4(x_6, x_4-\varepsilon_3, x_1+\varepsilon_3) \nn  \\
\cT_{\varepsilon_3} \act \widehat\varphi_3(x_5, x_2, x_6) &:=& \, \widehat\varphi_3(x_5, x_2, x_6).
\ees
The same field transformation can be expressed in a more explicit way (without star product) in the group representation, by group Fourier transform, as follows:  
\bes \label{VertexTranslation}
\cT_{\varepsilon_3} \act \varphi_1(g_1, g_2, g_3) &:=& \e_{g_1^\inv g_3}(\varepsilon_3) \, \varphi_1(g_1, g_2, g_3) \nn \\
\cT_{\varepsilon_3} \act \varphi_2(g_3, g_4, g_5) &:=&\e_{g_3^\inv g_4}(\varepsilon_3) \, \varphi_2(g_3, g_4, g_5) \nn \\
\cT_{\varepsilon_3} \act \varphi_4(g_6, g_4, g_1) &:=&\e_{g_4^\inv g_1}(\varepsilon_3) \, \varphi_4(g_6, g_4, g_1) \nn  \\
\cT_{\varepsilon_3} \act \varphi_3(g_5, g_2, g_6) &:=&\varphi_3(g_5, g_2, g_6).
\ees
The transformation is immediately extended to the complex conjugated fields $\overline{\vphi}_{\ell}$ by requiring consistency with complex conjugacy, using the property of the plane waves that $\overline{\e_g}(\varepsilon) = \e_{g^\inv}(\varepsilon)$.  

We see at first glance in (\ref{VertexTranslation1}) the geometric meaning of this transformation as a vertex translation, by the way it affects 
the arguments of the dual fields interpreted as edge vectors. When translanting a vertex of the triangle, one translates the two edges sharing this vertex; each edge is translated in an opposite way due to the orientation of the edges. If the vertex is shared by many edges, all these edges are translated accordingly, while taking into account their orientation. The gauge covariance of $\cT$ is manifest in both representations: in the metric representation, this is because the shift of the edge-variables preserves the closure of each triangle; in the group representation, this is because the arguments $g_j^{-1}g_k$ of the plane-waves are gauge invariant. 

Let us show that these field transformations leave the GFT action invariant. 
In fact, they leave invariant the field {\it polynomials} in this action, even before integration over the field variables. 
We check this in the group representation. 
Following the definition  (\ref{act}) of the translation algebra on a tensor product of fields, the action of the transformation 
$\cT_{\varepsilon_3}$ on the interaction polynomials
\[
\vphi_1(g_1, g_2, g_3) \vphi_2(g_3, g_4, g_5) \vphi_3(g_5, g_2, g_6) \vphi_4(g_6, g_4, g_1) 
\]
results in an overall multiplication by the phase:
\beq \label{trivialphase}
(\e_{g_1^\inv g_3} \star \e_{g_3^\inv g_4} \star \e_{g_4^\inv g_1})(\varepsilon_3) = \e_{g_1^\inv g_3 g_3^\inv g_4 g_4^\inv g_1}(\varepsilon_3) = 1.
\eeq
The interaction term is therefore invariant. Clearly, the ordering of fields and field arguments  is crucial. 
The kinetic term is also invariant, since $\cT_{\epsilon_3}$ acts on the field polynomials
$\varphi_\ell(g_1, g_2, g_3) \overline{\varphi}_\ell(g_1, g_2, g_3)$, for instance when $\ell = 1$,  by multiplication by the plane wave:
\[
(\e_{g_1^\inv g_3} \star \e_{g_3^\inv g_1})(\varepsilon_3) = (\e_{g_1^\inv g_3 g_3^\inv g_1})(\varepsilon_3) = 1.
\]
Thus, we have shown that the transformation generated by translation of the vertex $v_3$ is a symmetry of the GFT action. 
We can show similarly the invariance of the action under translations of the three other vertices $v_1, v_2$ and $v_4$ of the tetrahedron. 

Before discussing further the meaning of the symmetry in the next section, let us point out that the four symmetry generators are not all independent 
-- in other words, the symmetry is {\sl reducible}. In fact, there is a global translation of the four vertices of the tetrahedron under which the fields  transform trivially. Geometrically, this corresponds to the rather trivial fact that the geometry of a Euclidean triangle 
is invariant under a global translation of its vertices. 
Such a global translation is defined by choosing an order for the  vertices of each triangle. For example, choosing the order $v_3, v_4, v_2$ for the vertices of the triangle $\ell=1$, a global translation acts on $\varphi_1$ as
\[
\varphi_1 (g_1, g_2, g_3) \mapsto \e_{g_1^\inv g_3}(\varepsilon) \star \e_{g_2^\inv g_3}(-\varepsilon) \star \e_{g_1^\inv g_2}(-\varepsilon) \varphi_1 (g_1, g_2, g_3) = \varphi_1 (g_1, g_2, g_3).
\]

\subsection{Invariance of the vertex and diffeomorphisms}

We now want to show how the field symmetry (\ref{VertexTranslation}), and more specifically the invariance of the vertex function, 
tie together various notions of (discrete residual of) diffeomorphisms studied in the literature. To do so, we probe the meaning  of such  invariance in the different GFT representations. This picture will be completed in the next section, when we will discuss the invariance of the GFT Feynman amplitudes and n-point functions. 
  
\medskip

(i)  {\bf Metric representation.} 
The vertex function is given by the formula (\ref{Feynmanfunctions2}):
\beq
V(x_i^\ell, x_i^{\ell'}) = \int \prod_{\ell=1}^4 [\extd h_{\ell}]\, \prod_{i=1}^6 ( \delta_{\minus x^{\ell}_i} \star\, \e_{h_\ell h^{-1}_{\ell'}})(x^{\ell'}_i)
\eeq
Fixing the ordering of the variables to the one defined by the interaction polynomials, this function can be lifted to the group Fourier dual of a
tensor product in $\C(\SO(3))^{\otimes 12}$ invariant under the (non-commutative) translation (\ref{VertexTranslation0}). 
As we have already emphasized,  this transformation is geometrically interpreted as a translation of a vertex of the tetrahedron patterned by the interaction. 
More precisely, the function $V(x_i^\ell, x_i^{\ell'})$ imposes the variables $x_i^\ell$, interpreted as edge-vectors expressed in different frames, to match the metric of a Euclidean tetrahedron. 
The symmetry expresses the invariance of the matching condition under a translation of each of the vertices in an embedding of this  tetrahedron in $\R^3$.  
This is also how the action of discrete residual of diffeomorphisms is encoded in 3d Regge calculus. 

\medskip

\begin{figure}[h]
\begin{center}
\includegraphics[scale=.4]{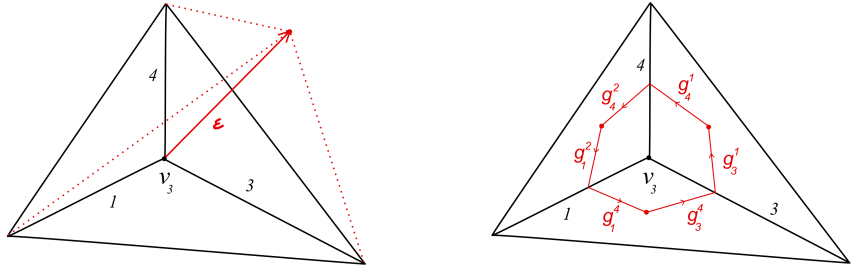} 
\caption{Vertex translation and trivial vertex holonomy} \label{translation-plouf}
\end{center}
\end{figure}

(ii) {\bf  Group representation}. 
The vertex function  is given by the formula (\ref{Feynmanfunctions}):
\beq \label{vertexfunc}
V(g_i^\ell, g_i^{\ell'})  = \int \prod_{\ell=1}^4 \extd h_{\ell} \prod_{i=1}^6 \delta((g^{\ell}_i)^{-1} h_\ell h^{-1}_{\ell'} g^{\ell'}_i)
\eeq
The invariance of this function under translations (\ref{VertexTranslation0})  of the vertex $v_3$ means that, for all $\epsilon_3 \in \su(2)$:
\beq
\e_{G_{v_3}}(\varepsilon_3) V(g_i^\ell, g_i^{\ell'}) = V(g_i^\ell, g_i^{\ell'})
\eeq
where the argument of the plane wave is:
\beq \label{Vertexhol}
G_{v_3} = (g^1_1)^{-1} g^1_3 (g^2_3)^{-1} g^2_4 (g^4_4)^{-1} g^4_1
\eeq
We thus see that translation invariance  reflects, in the group representation,  a conservation rule $G_{v_3} \!=\! 1$. 
Now, recall that the group field variables $g^\ell_i$ encode boundary holonomies, along paths connecting the center of each triangle $\ell$ to its edges. 
In the interaction term, they define a discrete connection living on the graph dual to the boundary triangulation of the tetrahedron, 
which has the topology of a 2-sphere. 
As illustrated on the right of Fig.  \ref{translation-plouf}, $G_{v_3}$ is the holonomy along a loop circling the vertex $v_3$ of the tetrahedron. 
Thus, the symmetry under translation of each vertex says the boundary connection is {\sl flat}. 
Imposing flatness $F=0$ of the boundary connection is precisely the role of the Hamiltonian and vector constraints, i.e. the canonical counterpart of diffeomorphisms, 
in (first order) 3d gravity \cite{karimalex}. Here, we see that the GFT symmetry results in such a constraint on the tetrahedral wave-function constructed from the GFT field. 


\medskip
 
(iii) {\bf Spin representation}. In the spin representation, obtained by Plancherel decomposition  
into $\SO(3)$ representations, the vertex  function takes the form of $\SO(3)$ 6j-symbols. 
This is a standard calculation, starting from the tensorial expression of $V$ in the $\SO(3)$ representations of spin $\{j_i^\ell\}$:
\beq \label{Vspin}
V^{j_{i}^\ell}_{m_i^\ell n_i^\ell} := \int \prod _{i} \extd g_i^\ell \extd g_i^{\ell'} \, V(g_i^\ell, g_i^{\ell'})  \prod _{i, \ell} D^{j_{i}^\ell}_{m_i^\ell n_i^\ell}(g_i^\ell) 
\eeq
where $D^j_{mn}(g)$ are the Wigner representation matrices. After a change of variables $g_i^\ell \to h^{-1}_\ell g_i^\ell$, 
the integration over the group elements  $h_\ell$ present in $V$ gives projectors onto the invariant tensors:
\beq \label{6j-1}
\int \extd h_\ell \, D^{j^\ell_1}(h_\ell)  D^{j^\ell_2}(h_\ell) D^{j^\ell_3}(h_\ell) = |i^\ell \rangle \, \langle i^\ell| ,
\eeq
where the intertwiners $i^\ell$ are the normalized Wigner 3j-symbols. 
The 6j-symbol, resulting from a contraction of four 3j-symbols that patterns a tetrahedron, shows up from the contraction of $V^{j_{i}}$ 
with the product of intertwiners $\prod_\ell i^\ell$ 
and the orthogonality of the Wigner matrices. 
\beq
\sum_{\{m_i^\ell\}} \prod_\ell i^\ell_{m_i^\ell}  \, V^{j_{i}}_{m_i^\ell n_i^\ell}  
= \prod_{i}  \delta_{n_i^\ell,\minus n_i^{\ell'}} \, \lower0.4in\hbox{\includegraphics[scale=.2]{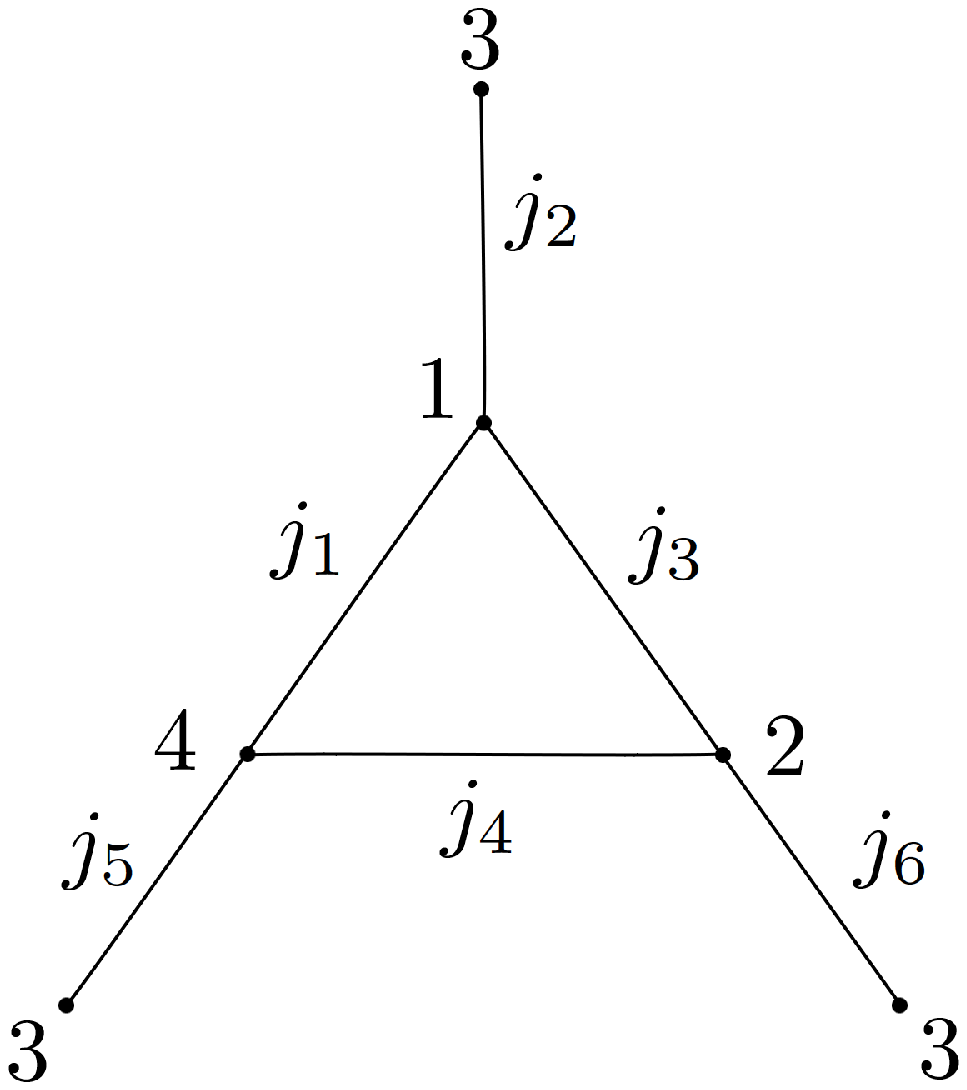}}
= \prod_{i} \delta_{n_i^\ell,\minus n_i^{\ell'}} \,
\left\{\begin{array}{ccc}
j_1 & j_2 & j_3 \\
j_4 &j_5& j_6
\end{array}
\right\} 
\eeq

There is a connection, also pointed out in \cite{ooguri, BC-WdW, EteraValentin}, between the flatness constraint described in (ii) and the
topological identities (Biedenhard-Elliot) satisfied by the 6j-symbol, which insures the formal invariance of the Ponzano-Regge spin foam model under refinement of the  triangulation. To see how the symmetry relates to such identities, let us make, within the integral (\ref{Vspin}), the (trivial) substitution:  
 \beq
V(g_i^\ell, g_i^{\ell'}) \to \int \extd k \, \e_{G_{v_3}}(k \varepsilon_3 k^{-1}) V(g_i^\ell, g_i^{\ell'})
\eeq
$G_3$ is the vertex holonomy given by (\ref{Vertexhol}); the factor in front of $V$ is the evaluation of a central function whose Plancherel decomposition is:
\beq
 \int \extd k \, \e_g(k \varepsilon_3 k^{-1}) = \sum_{j} \chi^j(g) \widehat{\chi}^j(\varepsilon_3) 
 \eeq
where $\chi^j$ is the $\SO(3)$ character in the spin $j$ representation and  $\widehat{\chi}^j\!\!=\!\!\int \extd g \chi^j(g)\e_g$
is its group Fourier transform\footnote{Explicitely, 
$\widehat{\chi}^j(\varepsilon)\!=\! J_{d_j}(|\varepsilon|)/|\varepsilon|$,  where $J_{d_j}$ is the Bessel function of the first kind associated to the integer $d_j\!:=\! 2j+1$,  peaked on the value $|\varepsilon|\!=\!d_j$.}. We also decompose into characters the three delta functions  in the expression of $V(g_i^\ell, g_i^{\ell'})$ (see Equ. (\ref{vertexfunc})) associated to the edges $i \!=\! 1,3,4$ sharing the vertex $v_3$, with the Plancherel formula $\delta(g) \!=\! \sum_k d_k \chi^k(g)$.
We thus obtain an expression in terms of the spins $j_i^\ell$ and a sum over four additional  spins $k_i$ and $j$. 
Elementary recoupling theory then shows that:
\bes \label{6j-2}
\sum_{\{m_i^\ell\}} \prod_\ell i^\ell_{m_i^\ell}  \, V^{j_{i}}_{m_i^\ell n_i^\ell}  
 &=& 
\sum_{k_i, j } d_{k_1} d_{k_3} d_{k_4} d_j   \widehat{\chi}^j(\varepsilon_3) \lower0.4in\hbox{\includegraphics[scale=.2]{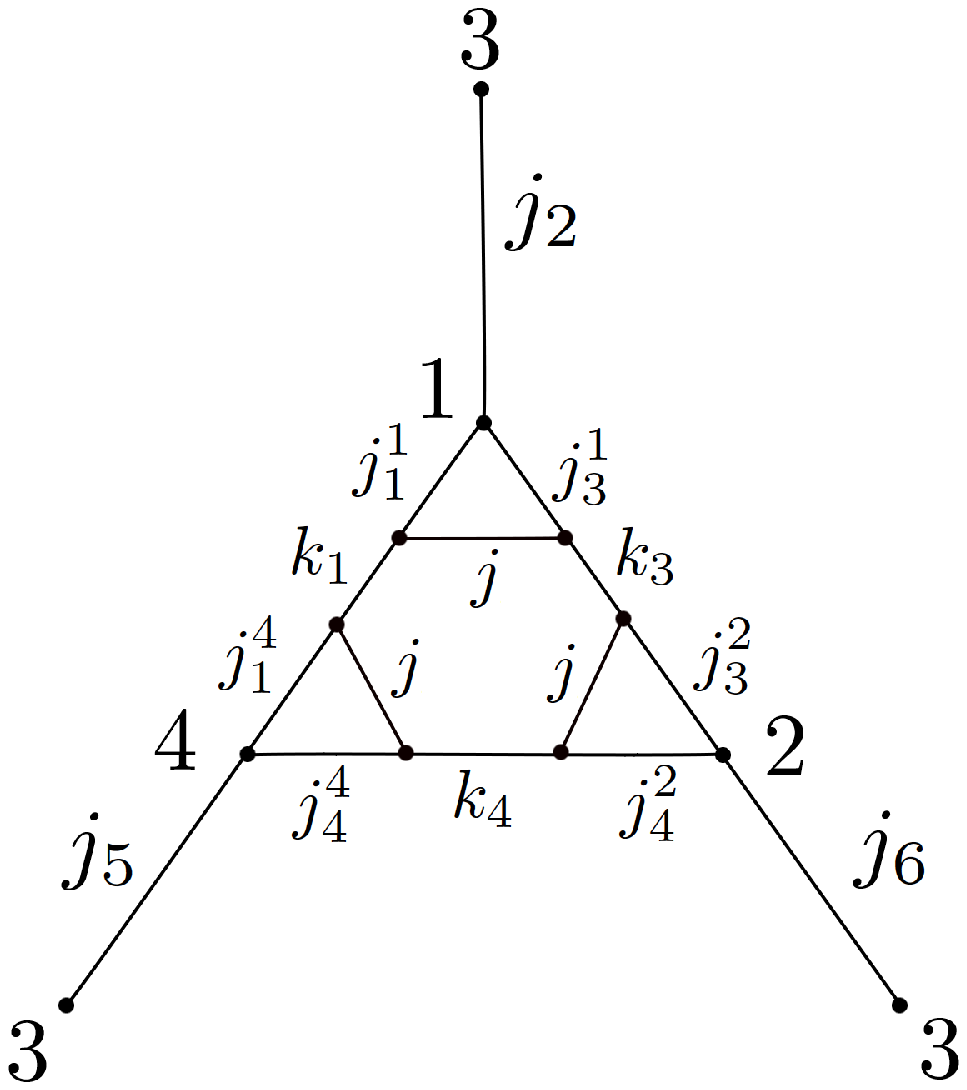}} \nn \\
&=&  \sum_{k_i, j } d_{k_1} d_{k_3} d_{k_4} d_j  \widehat{\chi}^j(\varepsilon_3)  
\left\{\begin{array}{ccc}
j_1 & j_2 & j_3 \\
j &k_1& k_3
\end{array}
\right\} 
\left\{\begin{array}{ccc}
j_1 & j_5 & j_4 \\
j &k_4& k_1
\end{array}
\right\} 
\left\{\begin{array}{ccc}
j_3 & j_6 & j_4 \\
j &k_3& k_4
\end{array}
\right\} 
\left\{\begin{array}{ccc}
k_1 & k_3 & j_2 \\
k_4 &j_5& j_6
\end{array}
\right\} 
\ees
Comparing (\ref{6j-1}) and (\ref{6j-2}), we obtain the following identities: 
\be
 \forall \varepsilon, \qquad \left\{\begin{array}{ccc}
j_1 & j_2 & j_3 \\
j_4 &j_5& j_6
\end{array}
\right\}  = \sum_{k_i, j } d_{k_1} d_{k_3} d_{k_4} d_j \, \widehat{\chi}^j(\varepsilon)  \, 
 \left\{\begin{array}{ccc}
j_1 & j_2 & j_3 \\
j &k_1& k_3
\end{array}
\right\} 
\left\{\begin{array}{ccc}
j_1 & j_5 & j_4 \\
j &k_4& k_1
\end{array}
\right\} 
\left\{\begin{array}{ccc}
j_3 & j_6 & j_4 \\
j &k_3& k_4
\end{array}
\right\} 
\left\{\begin{array}{ccc}
k_1 & k_3 & j_2 \\
k_4 &j_5& j_6
\end{array}
\right\} 
\ee

In turn, these identities imply recursion relations for the same 6j-symbols (see for e.g \cite{EteraValentin, Maite}), interpreted as discrete versions of the Wheeler-de-Witt equation \cite{BC-WdW}. More generally, we expect that our type of analysis, based on GFT symmetries, can give a systematic way, also for gravity models in higher dimension,  to derive algebraic identities of the spin foam quantum amplitude from the study of the GFT symmetries.

 \medskip

This gives a clear interpretation of the symmetry in the various representations of the GFT, which matches what we expect from the action of diffeomorphisms in discrete approaches. Thus, in the metric representation, the symmetry encodes  the invariance under (non-commutative) translation of the four vertices of the tetrahedron. In the group picture, they encode the flatness of the discrete boundary connection, which is the Wheeler-DeWitt constraint in connection variables, and thus the canonical diffeomorphism constraints. In the spin picture, they encode recursion relations for the fundamental spin foam amplitudes (6j-symbols) and their behavior under coarse-graining.

\subsection{GFT with vertex variables}\label{sec:vertex-variables}


We have seen that the invariance of the vertex function reflects some conservation rules  for the holonomies $G_v$ 
along loops surrounding the vertices of the  tetrahedron patterned by the interaction. 
These conservation rules can be made  manifest by integrating out the gauge group element $h_\ell$ in the vertex function.
Using three of the six delta functions in (\ref{Feynmanfunctions}) to integrate three of the four integration variables $h_\ell$, we obtain:
\begin{align*}
V(g_i^\ell, g_i^{\ell'}) \!&=\! \int \prod_{\ell=1}^4 \extd h_\ell \, \prod_{i=1}^6 \delta((g^{\ell}_i)\mone h_\ell h\mone_{\ell'} g^{\ell'}_i) \nn\\
\!&=\! \int[ \extd h_4 ]\,
\delta((g_{4}^2)\mone g_5^2(g_5^3)\mone g_{6}^3 (g_6^4)\mone g_{4}^4)
\delta((g_1^1)\mone g_{2}^1 (g_2^{3})\mone g_6^3 (g_6^4)\mone g_{1}^4 )
\delta((g_1^1)\mone g_{3}^1 (g_{3}^2)\mone g_4^2 (g_4^4)\mone g_1^4) .
\end{align*}
Thanks to the normalization of the Haar measure, this simply gives:
\beq 
V(g, g') = \delta(G_{v_1}) \delta(G_{v_2}) \delta(G_{v_3}). 
\eeq
Note that the fourth constraint $G_{v_4}\!=\!1$  is a consequence of the other three, 
due to the dependence relation $ G_{v_4} \mone(g_3g_{3'}\mone)  (G_{v_2} G_{v_3} G_{v_1})(g_{3'}g_3\mone) =1 $ between the four vertex holonomies. This is the counterpart of the reducibility of the translation symmetry studied in the previous section. The dependence relation can be  easily  understood as a discrete Bianchi identity for the boundary connection on the boundary surface of the tetrahedron. 

This form of the vertex function suggests yet another representation of GFT in terms of {\sl vertex} variables $v_i\!\in\! \su(2)$ instead of edge vectors  $x_i$. These vertex variables, which are the generators of the translation symmetry,  are introduced  by plane-wave expansion $\delta(G_{v_i}) \!=\!\int \extd^3 v_i  \,\e_{G_{v_i}}(v_i)$ of the delta-functions on the group. Writing each of these plane-waves as a cubic term,  for e.g 
\beq
\e_{G_{v_1}}= \e_{(g_{4}^2)^\inv g_5^2} \star \e_{(g_5^3)^\inv g_{6}^3} \star \e_{(g_6^4)^\inv g_{4}^4},
\eeq
suggests to recast the GFT interaction in terms of new fields defined by  Fourier transform
\bes
\hpsi_1(v_2, v_3, v_4) := \int \extd g_1 \extd g_2 \extd g_3 \e_{g_1^\inv g_2}(v_2) \e_{g_1^\inv g_3}(v_3) \e_{g_2^\inv g_3}(v_4) \, \vphi_1(g_1, g_2, g_3) \nn \\
\hpsi_2(v_1, v_3, v_4):= \int \extd g_3 \extd g_4 \extd g_5 \e_{g_4^\inv g_5}(v_1) \e_{g_3^\inv g_4}(v_3) \e_{g_3^\inv g_4}(v_4)\,  \vphi_2(g_3, g_4, g_5) \nn \\
\hpsi_3(v_1, v_2, v_4):=\int \extd g_5 \extd g_2 \extd g_6 \e_{g_5^\inv g_6}(v_1) \e_{g_2^\inv g_6}(v_2) \e_{g_4^\inv g_2}(v_4) \, \vphi_3(g_5, g_2, g_6) \nn \\
\hpsi_4(v_1, v_2, v_3):= \int \extd g_6 \extd g_4 \extd g_1 \e_{g_6^\inv g_4}(v_1) \e_{g_6^\inv g_1}(v_2) \e_{g_4^\inv g_1}(v_3) \, \vphi_4(g_6, g_4, g_1)
\ees
In terms of these new fields, well-defined on gauge invariant fields $\cP\acts \vphi_\ell$, the combinatorics of the GFT interaction patterns now the combinatorics of the four {\it vertices} $v_1 \cdots v_4$ in the four triangles in a tetrahedron, with a consequent change in the diagrammatic representation. Using the short notation $\hpsi^{123} \!:=\! \hpsi(v_1, v_2, v_3)$, we in fact obtain:
\beq
S_{int}[\hpsi] = \lambda \int  [\extd^3 v_i]^3\, \hpsi_1^{234} \star \hpsi_2^{134} \star \hpsi_3^{124} \star \hpsi_4^{123} +
\lambda \int  [\extd^3 v_i]^3\,\widehat{\overline{\psi}}_4^{\lower0.1in \hbox{\footnotesize 321}} \star \widehat{\overline{\psi}}_3^{\lower0.1in \hbox{\footnotesize421}} \star \widehat{\overline{\psi}}_2^{\lower0.1in \hbox{\footnotesize431}} \star \widehat{\overline{\psi}}_1^{\lower0.1in \hbox{\footnotesize 432}}.
\eeq
Here the $\star$-product relates repeated indices as $\hpsi^i \star \hpsi^i \star \hpsi^i \!=\! (\hpsi\star \hpsi \star \hpsi)(v_i)$. 
In each integral, the integration is over three variables $v_1, v_2, v_3$ only: the value of $v_4$ is pure gauge, fixed to an arbitrary value by global translation. 
A similar analysis can be performed for the kinetic term. In terms of the  fields $\hpsi_\ell$, it is given by 
\beq
S_{kin}[\hpsi] = \sum_\ell\int [\extd^3 v_i]^2\, \hpsi_\ell^{123}  \star \widehat{\overline{\psi}}_\ell^{\lower0.1in \hbox{\footnotesize123}}
\eeq
where the integration  is over two variables $v_1, v_2$ only, the value of $v_3$ being arbitrary fixed using translation invariance.

We do not analyze further this reformulation of the model in terms of vertex variables, in this paper. 
We however believe that it will be relevant in many respects. First, the properties of the GFT following from our diffeomorphism transformations should be simpler to analyze in this formulation, since it is in fact on the vertex variables that these transformations act naturally and in the simplest way. 
Second, the formulation of the GFT amplitudes in terms of vertex variables should simplify the analysis of their divergences,  
which are known to be located on 3-bubbles of GFT diagrams, namely at the vertices of the  triangulation, in addition to a global dependence on the overall topology of the diagrams \cite{laurentdiffeo,FGO,vincentmatteo,ValentinMatteo,PR1}.

\section{From GFT to simplicial gravity symmetries}
\label{GFTvsSFsymmetries}

We have seen in Sec. \ref{introGFT} that the amplitude of a Feynman GFT diagram, in the metric representation, gives the simplicial path integral form of the Ponzano-Regge model. In this section, we investigate how the GFT symmetry described above relates to the discrete residual action of diffeomorphisms in this model \cite{laurentdiffeo}.  

\subsection{Diffeomorphisms in simplicial path integrals}

The amplitude of a closed Feynman GFT diagram $\cG$, in the metric representation, takes the form:
\beq \label{pathintegral}
Z_{\Delta} = \int \prod_l \extd h_l \prod_e \extd^3 x_e \,\, e^{i S_{\Delta}(x_e, h_l)},
\eeq
where $\Delta$ is the simplicial complex dual to $\cG$ and $S(x_e, h_l)$  is the discrete 3d gravity action:
\beq \label{weight}
e^{i S_{\Delta}(x_e, h_l)} := e^{i \sum_e \Tr  x_e H_e} = \prod_e \e_{H_e}(x_e).
\eeq
The variables of this action are a discrete metric $\{x_e \}_{e \in \Delta}$ on the edges of the triangulation and a discrete connection 
$\{h_l\}_{l \in \cG}$ on the lines of $\cG$. The group element 
$H_e ={\overrightarrow{\prod}}_{l \in \partial f_e} \, h_l$
is the holonomy along the boundary of the face $f_e$ of $\cG$ dual to $e$, computed from a given reference vertex in $\partial f_e$. 
In the case of open diagrams and in the presence of boundary data $f(x)$, the integrand is obtained by taking the $\star$-product $f \star e^{i S_{\Delta}}$, with respect to the boundary variables $\{x_e\}_{e \in \partial \Delta}$. 

The action $S_{\Delta}(x_e, h_l)$  is a discrete version of the continuum action for first order 3d gravity:
\beq \label{action}
S(B, A) \!=\! \int \Tr \, B \wedge F,
\eeq
where $B$ is the triad frame field and $F$ is the curvature of the connection $A$. We recall in Appendix \ref{app} the local Poincar\'e symmetry of the continuum theory, namely the $\SO(3)$ gauge invariance and translation symmetry:  
\beq  \label{translation}
\begin{array}{l}
B \to B+ d_A \phi\\
A\to A 
\end{array}\quad 
\left|\quad  \begin{array}{l}
B \to [ B, X] \\
A\to A+ dX+ [ A, X], \quad F\to F+ [F,X],
\end{array}\right.
\eeq 
with both $X$ and $\phi$ scalars with value in $\su(2)$. The  translation symmetry, typical of BF-type theories, is due to the Bianchi identity $\extd_A F \!=\! 0$. 
As we show in Appendix, the action of diffeomorphisms in 3d gravity is classically equivalent to (a combination of gauge transformations and) translation of the frame field. 

The action $S_{\Delta}(x_e, h_l)$ enjoys a discrete version of these symmetries \cite{laurentdiffeo}.  
It can moreover be shown that, whenever $\Delta$ triangulates a 3-manifold, the discrete residual  of translation invariance, 
and hence of diffeomorphism invariance, in the discrete path integral (\ref{pathintegral}), is partially\footnote{For a finer analysis of the divergences of the Ponzano-Regge model, see \cite{ValentinMatteo}.} 
responsible for the large-spin divergences in the Ponzano-Regge model \cite{laurentdiffeo, PR1}. 

The discretization of the gauge transformations follows the usual lattice gauge theory techniques. The generators $\Lambda_v$ are labelled by vertices of the  GFT graph $\cG$ - equivalently by tetrahedra in the triangulation $\Delta$. 
The holonomy $h_l$ on the oriented lines of $\cG$ are transformed as  $h_l \to \Lambda_{v_s} \, h_l \, \Lambda_{v_t}\mone$, where $v_s, v_t$ denote the source and target vertices of the line $l$. This means in particular that the holonomy $H_e$ around the boundary of a face $f_e$  is transformed as $H_e \to \Lambda_e \, H_e \, \Lambda_e\mone$, where $\Lambda_e$ is the generator associated to the reference vertex in $\partial f_e$ from which the holonomy is computed. 
The metric variable $x_e$ transforms as $ x_e \to \Lambda_e \, x_e \, \Lambda_e\mone$.  
Such a gauge transformation, under which the action is clearly invariant, corresponds to a rotation of the reference frame of the $e$. 

The discrete residual of translation invariance is due to a discrete analogue of the Bianchi identity  satisfied by the curvature elements $H_e$. 
In terms of the GFT diagram and its dual simplicial complex $\Delta$, this can be understood as follows. 
Given a vertex $v \in \Delta$, the set of GFT faces $f_e$ dual to the edges $e \supset v$ meeting at $v$ defines a cellular decomposition of a surface $L_v$, called  the {\sl link} of the vertex $v$. In GFT language, the link of a vertex is the boundary of a 3-dimensional `bubble' of the diagram. 
Whenever the simplicial complex $\Delta$  defines a triangulated manifold (as opposed to a pseudo-manifold), the link of every vertex 
has the topology of a 2-sphere.
Then for any ordering of the edges $e \supset v$ meeting at $v$, the curvature elements $H_e$ satisfy a closure relation of the type 
\beq \label{Bianchi}
\overrightarrow{\prod_{e \supset v}} (k_v^e)^{-1}\, H_e \, k_v^e  = 1
\eeq
for some group-valued  functions $k_v^e:= k_v^e(h_l)$ of the variables $h_l$ on the links $l$ of $L_v$. 
The group elements $k_v^e$ are interpreted as the parallel transport along paths between a fixed vertex in $L_v$ to the reference vertex of each face $f_e$ from which the holonomy $ H_e$  computed. We have assumed here that the orientation of the faces $f_e$ all agree with a fixed orientation of the sphere $L_v$; if the orientation of $f_e$ is reversed, $H_e^{-1}$ should appear in place of $H_e$. Note that no such closure identity holds when the $L_v$ has a higher genus topology. 

The idea of the works \cite{laurentdiffeo, PR1} was to use the identities (\ref{Bianchi}) to prove a (commutative) discrete analogue of translation symmetry 
for the discrete action $S_{\Delta} = \sum _e \Tr x_e H_e$. 
To do so, the identity (\ref{Bianchi}) is first written in terms of the projections $P_e := \Tr H_e\vec{\tau}$ of the curvature elements onto the Lie algebra:
\beq
\sum_{e \supset v} (k_v^e)^{-1}  (U^v_e P_e + [\Omega^v_e, P_e]) k_v^e= 0,
\eeq
where the scalar $U^v_e$ and Lie algebra elements  $\Omega^v_e$ are certain (complicated) functions of the $P_e$'s obtained from the Campbell-Hausdorff formula \cite{laurentdiffeo}. 
This leads to the invariance of $S_{\Delta}$ under the following transformation, generated by $\epsilon_v \in \su(2)$:
\beq \label{commdiscretediff}
x_e \, \mapsto \, x_e +   U^v_e \varepsilon_v^e  - [\Omega^v_e,  \varepsilon_v^e ], \quad \textrm{ with }\varepsilon^e_v(\epsilon_v) = k^v_e\epsilon_v (k^v_e)^{-1}.
\eeq
 Note that, if $\epsilon_v$ is interpreted as a translation vector in the reference vertex frame of $L_v$, 
$\varepsilon^e_v$ encodes the same translation vector parallel-transported  in the reference frame of $f_e$.  
The transformation (\ref{commdiscretediff}) is a discrete analogue of the translation symmetry (\ref{translation}).

In the next section, we show that the discrete Bianchi identity (\ref{Bianchi})  can be related to a vertex translation symmetry in a direct way -- that is, without any projection to the Lie algebra --  provided one takes into account the {\sl non-commutativity} of the translation algebra  studied in Sec. \ref{sec:dsu2}. 
This will clarify the relationship between the GFT symmetry  and the discrete Bianchi identities leading to diffeomorphism invariance in the simplicial path integrals.

\subsection{Simplicial diffeomorphisms as quantum group symmetries}


To see how the discrete Bianchi identities are tied to the invariance under noncommutative vertex translation defined in Sec. \ref{sec:sym}, 
let us fix the value $x_e$ of the metric in the exponential of the action (\ref{weight}), for all edges $e$ which do not touch the vertex $v$. 
This defines a function of the remaining $n_v$ variables in $\su(2)$, labelled by the $n_v$ edges $e\supset v$ sharing  $v$. 
Choosing an ordering of these edges as in (\ref{Bianchi}), one can lift this function to an element of the tensor product $\cC(\SO(3))^{\otimes_{e \supset v}}$ of $n_v$ copies of $\cC(\SO(3))$. 

Let us now act with the non-commutative translation 
\beq
x_e \to x_e + \varepsilon_v^e(\epsilon_v), \qquad \varepsilon^e_v(\epsilon_v) = k^v_e\epsilon_v (k^v_e)^{-1}
\eeq
shifting the metric  of the edges sharing $v$ by the variables $\varepsilon^e_v$ defined as in (\ref{commdiscretediff}).  The  group elements $k^v_e$ parallel transport the frame of a fixed vertex in $L_v$, and that of the reference vertex of the face $f_e$ from which the holonomy $H_e$ is computed. 
Upon such a  translation,  the function (\ref{weight}) gets transformed into a $\star$-product of functions of $\epsilon_v$: 
\beq \label{vertexsymmetry}
\prod_e e^{i \Tr x_e H_e} \,\, \mapsto \,\, \underset{e \supset v}{\overrightarrow{\bigstar}}\prod_e e^{i \Tr (x_e + \varepsilon_v^e) H_e}(\epsilon_v) .
\eeq
Using the rule (\ref{star})  for the $\star$-product of plane-waves, we see that such  a non-commutative translation acts on the action term by multiplication by the plane wave:
\beq \label{InvariantAction}
e^{i \Tr \,\left[ \epsilon_v  \left(\overrightarrow{\prod}_{e \supset v} (k_v^e)^{-1}\, H_e \, k_v^e\right)\right]} = 1 ,
\eeq
which is trivial due to the Bianchi identity (\ref{Bianchi}).

As we have seen in Sec. \ref{sec:sym}, both the GFT propagator and vertex functions, which the Feynman integrand (\ref{weight}) is built upon, 
are invariant under  vertex translation $x_e \to x_e + \epsilon_v$, for $e \supset v$. 
The generator $\epsilon_v$  is interpreted as a translation vector in a given frame. This is the frame associated to the reference point of the loop circling $v$, along which the conserved holonomy is computed. In  (\ref{trivialphase}), for example, this is the frame associated to the edge 1 of the tetrahedron patterns by the interaction\footnote{ 
As made clear using the covariance of the plane wave upon conjugacy, the same  translation expressed in a different  frame, say that of edge 3, is generated by $\epsilon'_v \!:=\! k \epsilon_v k^{-1}$, where $k:=g_3^{-1} g_1$ parallel transports one frame to another.}.

The transformation (\ref{vertexsymmetry}) has the same geometrical meaning: it corresponds to a vertex translation expressed in a given frame. 
This frame is the reference vertex frame of the link $L_v$. Indeed, recall from the calculation of the Feyman amplitudes in Sec. \ref{introGFT} that the variable $x_e$ present in the action term is the edge-metric in the reference frame of the GFT faces $f_e$ dual to the $e$.  
Using the parallel transports $k_e^v$, one could instead use variables $x^v_e$ labeling to the same edge-metric, but expressed
in the reference vertex frame of $L_v$. These are defined by $x_e= k_e^v x^v_e (k_e^v)^{-1}$.  Now, in this frame, a vertex translation acts as $x_e^v  \to x_e^v + \epsilon_v$. This amounts to act on the original variables $x_e$ by the `twisted' translation $x_e \to x_e + \varepsilon^e_v$, where $\varepsilon^e_v(\epsilon_v) = k^v_e\epsilon_v (k^v_e)^{-1}$. 

Thus, the equality (\ref{InvariantAction}), and hence the discrete Bianchi identity, express the invariance of the exponential of the action under the (quantum) GFT symmetry defined in the previous section. Note that, interestingly,  the analysis of the invariance under simplicial diffeomorphisms distinguishes the closed GFT diagrams $\cG$ which define a {\sl manifold} from those defining only a {\sl pseudo-manifold}. In fact, in the case of non-manifold graphs, the triangulation has vertices $v$ for which the link $L_v$ defines a surface with non-trivial topology. For such vertices, there is no analogue of the discrete Bianchi identity (\ref{Bianchi}): the invariance of the exponential of the action (\ref{weight}) under vertex translation is therefore broken. 

The goal of the next subsection is to illustrate how these rather geometric considerations can be understood in a purely algebraic way. 
We will show on a simple example how the use of  {\sl braiding}  techniques could give a systematic way to derive Bianchi identities from the GFT symmetry.


\subsection{Non-commutative translations, invariance of the GFT amplitudes and Bianchi identities}

As spelled out in Sec. \ref{introGFT}, the integrand of a GFT Feynman amplitude in  the metric representation is calculated by sticking together  propagator and vertex functions along  each loop $f_e$ of the diagram, using the $\star$-product. This gives a product loop amplitudes (see  \eqref{Feyn-integrand}):
\beq \label{integrand}
\prod_{f_e}\, \vec{\bigstar}_{j=0}^{N + 1} \, (\delta_{x_e^j} \star \e_{h_{jj+1}})(x_e^{j+1}) .
\eeq
The exponential of the discrete BF action (\ref{weight}) is then obtained by integrating, within each loop, over all metric variables $x_e^j$, save one $x_e:= x_e^0$. 
It was shown in the previous section that the GFT propagator and vertex define invariant functions under the non-commutative translation (\ref{VertexTranslation0}). 
The question we are asking is to which extent the translation invariance of the propagator and vertex functions induces the invariance (\ref{vertexsymmetry}), (\ref{InvariantAction}) of the action term. We will only sketch an answer here with a simple example, leaving the full proof to future work. 

Since the transformation is quantum symmetry, it is crucial, to answer the above question, to keep track of the ordering of the variables in the calculation of the Feynman integrand.  It is precisely to keep track of this ordering that the braided quantum field theory formalism \cite{oeckl} uses a perturbative expansion into {\sl braided} Feynman diagrams.
 
Note that, to study the behavior of  (\ref{integrand}) under the translation of a vertex $v$, it is enough to restrict the product to the set of loops $f_e$ such that $e \supset v$. This amounts to considering  the contribution of a subdiagram called `3-bubble' \cite{razvan}, which represents the vertex $v$. 
A 3-bubble, obtained by erasing all lines having strands of a given color, is a trivalent ribbon graph dual to the link $L_v$ of a vertex of the dual triangulation.  

 \begin{figure}[t]
\begin{center}
\label{diag 2points}
\includegraphics[scale=0.5]{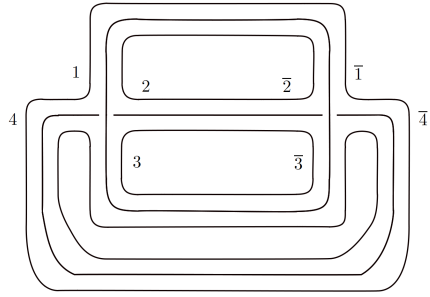} 
\caption{2 vertex Feynman diagram encoding the discretization of the sphere $S^3$.}
\end{center}
\end{figure}

The Fig. 6  shows the simplest GFT diagram of order two, dual to a triangulation of the sphere $S^3$ with two 3-simplices; 
the Fig. 7 shows the 3-bubble obtained by erasing all the lines having strands of color $4$. 
The 3-bubble can be drawn as a {\sl braided diagram}, on the left of Fig 8: each vertices are put beside each other, all legs up, in the lower part of the diagram, in a way that preserves the cyclic order of the legs on the plane; then the propagator strands, in the upper part of the diagram, connect the legs with each other. 
A convenient way to represent these vertices is as a product of three `cups' (see Fig 8): 
\beq
\lower0.2in\hbox{\includegraphics[scale=.4]{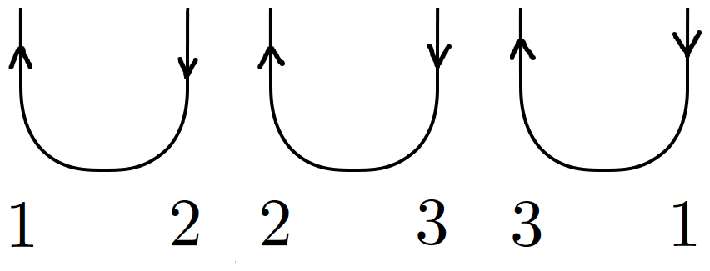}} 
\eeq

 \begin{figure}[t]
\begin{center}
\label{bubble}
\includegraphics[scale=0.5]{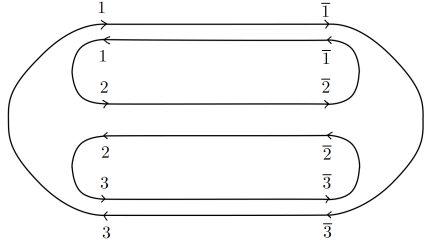} 
\caption{3-bubble dual to the vertex 4, obtained from Fig. 6 by erasing the strands related to color 4.}
\end{center}
\end{figure}


The Feynman rules to compute the contribution of the 3-bubble to the amplitude are easily read from (\ref{Feynmanfunctions2}). 
If one reabsorb the minus sign and group variables of the propagator into the vertex,  we get a contribution of each `cup'  
\lower0.15in\hbox{\includegraphics[scale=.3]{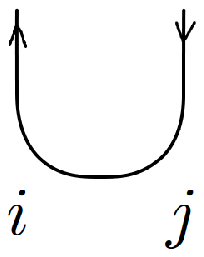}} given by 
\beq \label{cup}
\lower0.2in\hbox{\includegraphics[scale=.4]{cup.eps}} \,\, = \,\,  (\delta_{x^i_{ij}} \star \e_{h_i^\inv h_j})(x_{ij}^j)
\eeq
whereas the propagator strands are just non-commutative delta-functions $\delta_x(x')$.

Upon non-commutative translation $x \to x+\epsilon$, the `cup'  function of two variables $x_{ij}^i,x_{ij}^j$ gets transformed as 
\beq
\cT_\epsilon \,\acts \,\, \lower0.2in\hbox{\includegraphics[scale=.4]{cup.eps}} \,\, = \,\, 
\e_{h^\inv_i h_j}(\epsilon) \,\, \lower0.2in\hbox{\includegraphics[scale=.4]{cup.eps}} 
\eeq
This can be easily seen by group expansion of the non-commutative delta functions. 
One can then convince oneself that the translation invariance of the vertex function is then due to the following identities: 
\beq
\cT_\epsilon \,\, \acts \lower0.2in\hbox{\includegraphics[scale=.4]{3-cups.eps}}  = 
\e_{h^\inv_1 h_2 h_2^\inv h_3 h_3^\inv h_1}(\epsilon)\lower0.2in\hbox{\includegraphics[scale=.4]{3-cups.eps}}= \lower0.2in\hbox{\includegraphics[scale=.4]{3-cups.eps}}
\eeq
Hence, we see that, by construction, the lower part of the braided bubble diagram defines a translation invariant function of the metric variables.

Now, the contribution of the 3-bubble appears in the final integrand (\ref{integrand}) as a product of loops, as drawn on the right of  Fig. 8. 
\begin{figure}[t]
\begin{center}
\includegraphics[scale=.4]{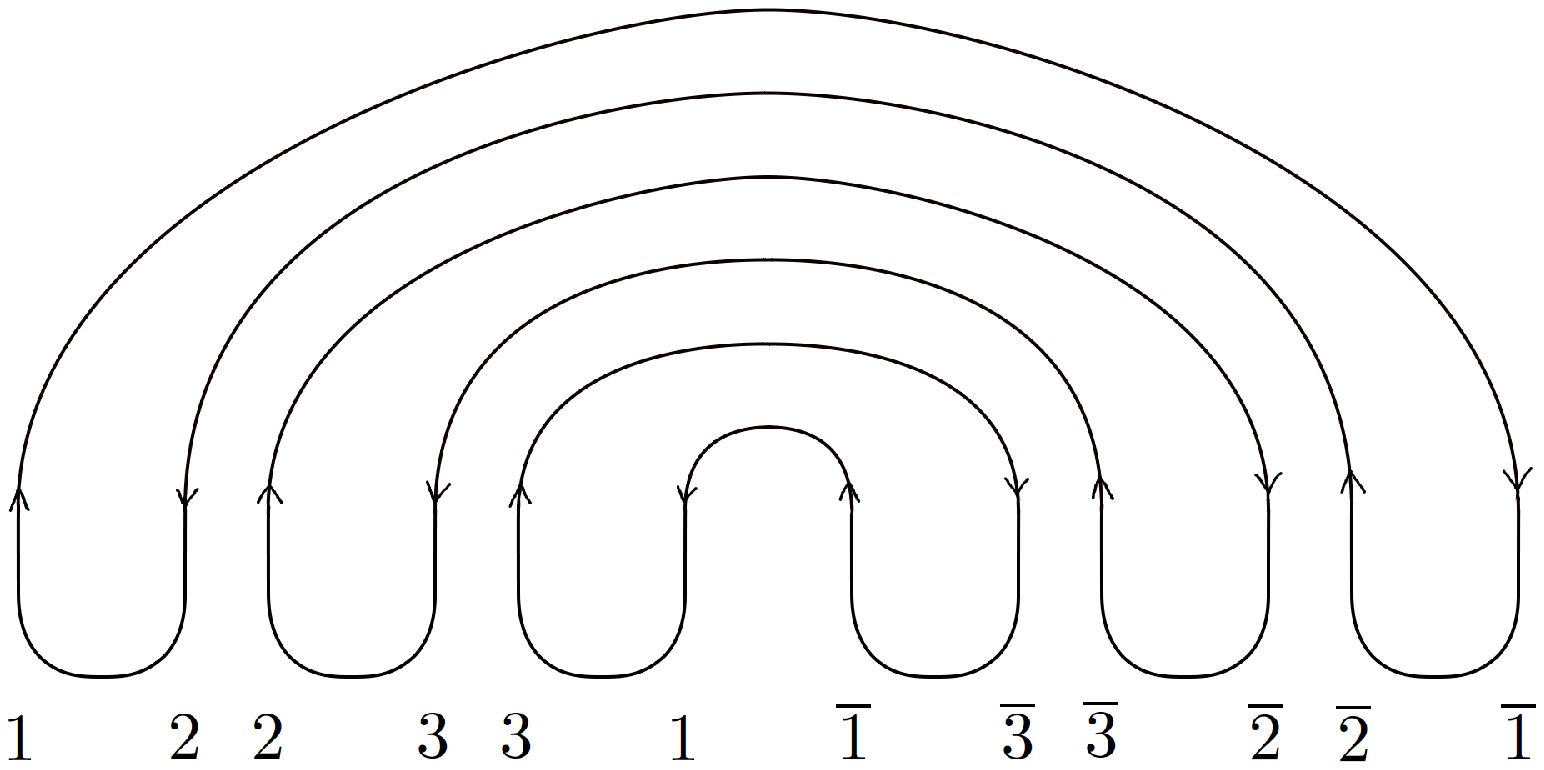}
\includegraphics[scale=.4]{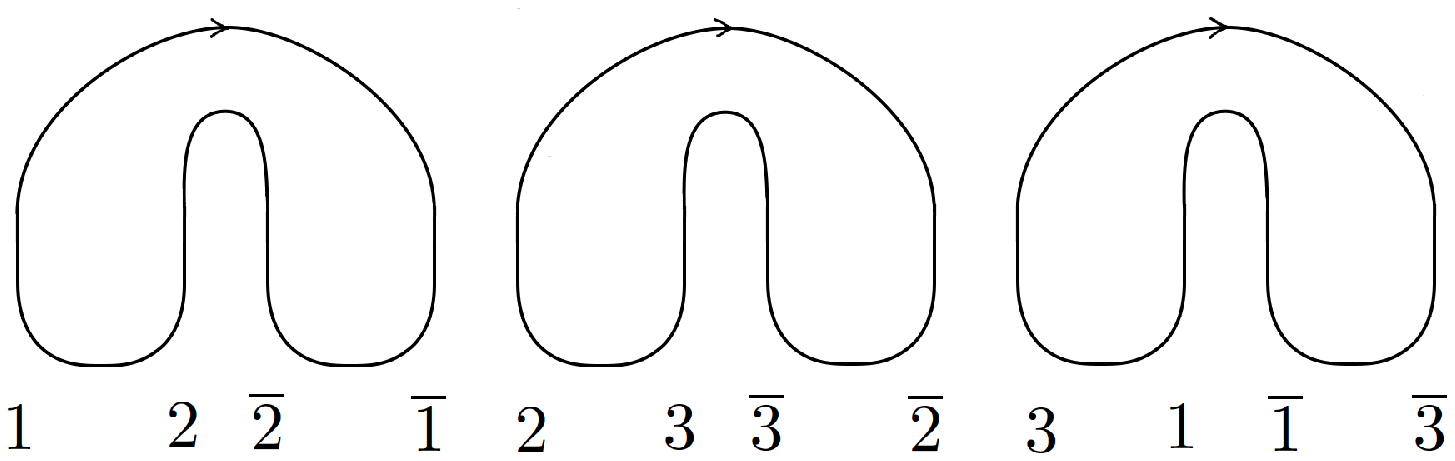}
\caption{3-bubble drawn (i) as a braided diagram and  (ii) as a product of loops.} \label{3-bubble}
\end{center}
\end{figure}
Going from the left to the right diagrams in Fig 8 by `separating the loops' induces re-ordering of the strands -- hence a re-ordering of the metric variables. 
In order to probe the behavior of (\ref{integrand}) under translation, the idea is to associate to a certain {\sl braiding map} to the separation of the loops, induced by the universal R-matrix of $\cD\SO(3)$ given in (\ref{braiding})\footnote{Let us stress that this braiding is merely a technical aid to keep track of the action of our quantum group symmetry on functions of several Lie algebra variables, and does not correspond here to any non-trivial braiding in the algebra of GFT fields, which we have chosen to be trivial, as in the standard GFT framework.}. As a direct calculation shows, swapping two cups (the right one above the left one) with the $\cD\SO(3)$ braiding gives:
\beq \label{swapcups}
\lower0.5in\hbox{\includegraphics[scale=.4]{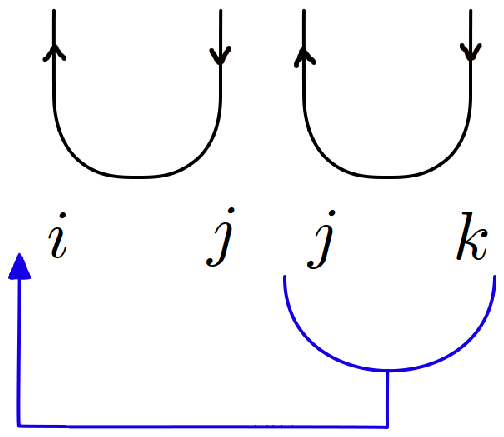}}=  
\lower0.2in\hbox{\includegraphics[scale=.4]{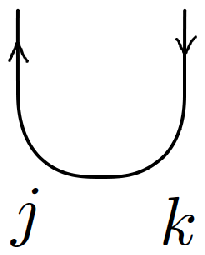}}  \, h_j^{\minus 1} h_k \acts \lower0.2in\hbox{\includegraphics[scale=.4]{cup.eps}} 
\eeq
where $h \acts \lower0.2in\hbox{\includegraphics[scale=.4]{cup.eps}}$ denotes the action of $h$ by conjugacy on the two variables of the cup: 
\beq
h \acts \lower0.2in\hbox{\includegraphics[scale=.4]{cup.eps}} :=( \delta_{h^\inv x_{ij}^i h} \star \e_{h_i^\inv h_j})(h^{\minus 1} x h).
\eeq
By construction, swapping the cups in this way intertwines the translation $\cT_{\epsilon}$. 
 
Let us now use this braiding to `separate the loops'. The loop $12\bar{2}\bar{1}$ is separated as follows:
\beq\nn
\includegraphics[scale=.4]{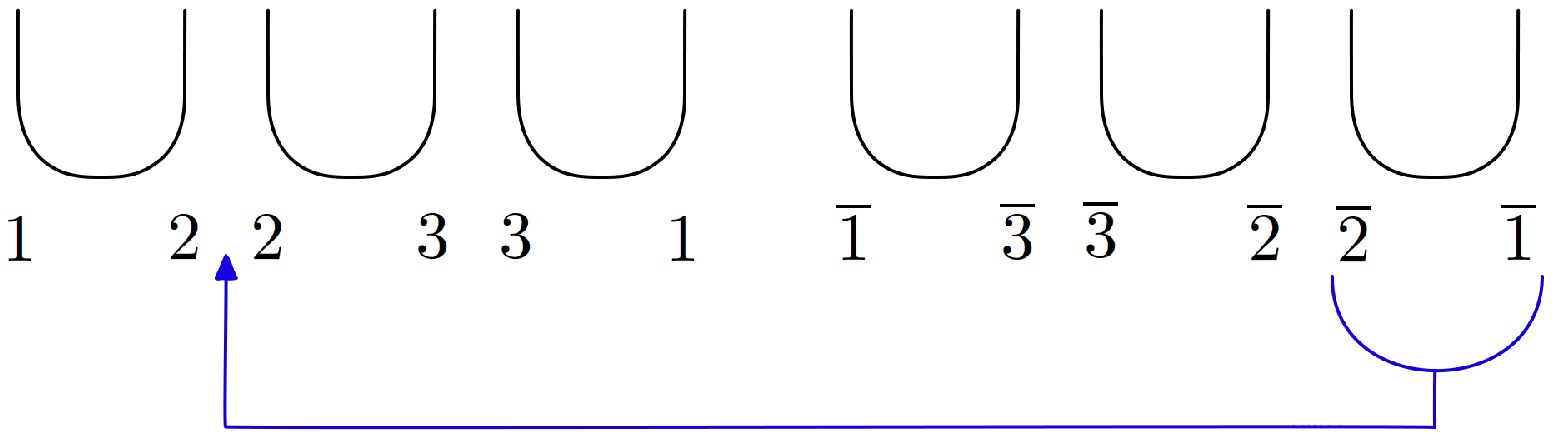} 
\eeq
The next step is to form the loop $23 \bar{3} \bar{2}$. 
\beq\nn
\includegraphics[scale=.4]{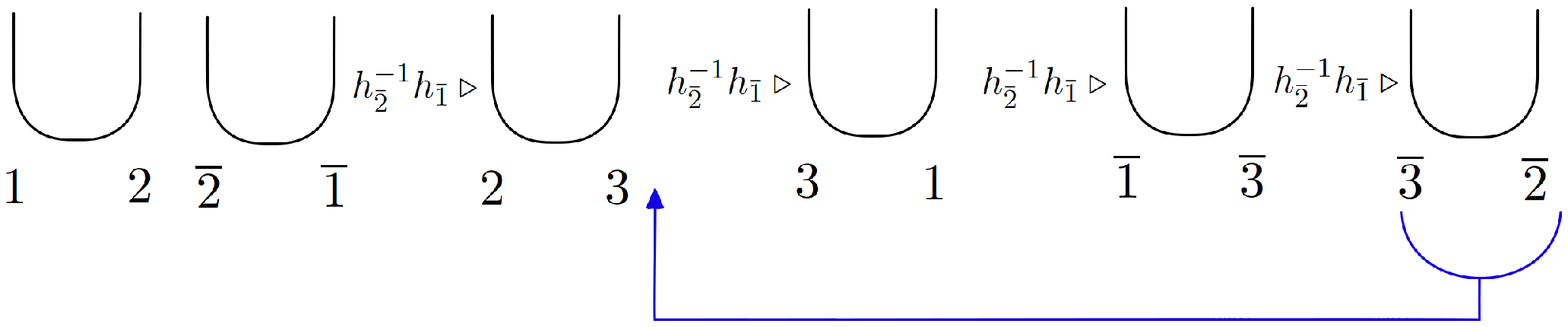} 
\eeq
Hence, using the $\cD\SO(3)$ braiding to separate the loops finally gives {\sl twisted} product of loops: 
\beq \label{twistedloops}
\lower0.3in\hbox{\includegraphics[scale=.4]{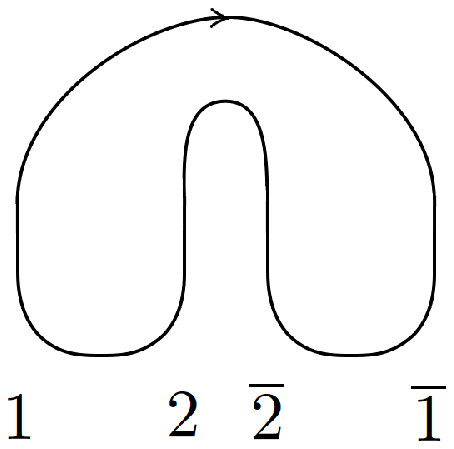}}\quad  h^{-1}_{\bar{2}} h_{\bar{1}} \acts \lower0.3in\hbox{\includegraphics[scale=.4]{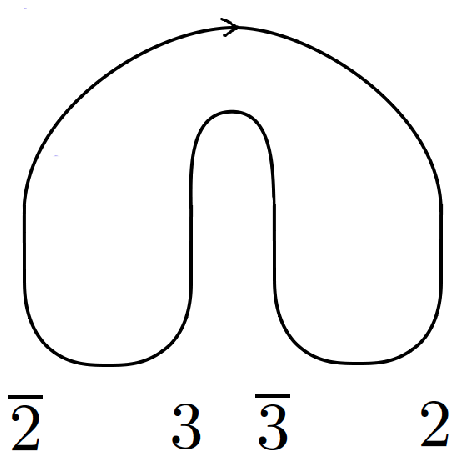}} \quad h_{\bar{3}}^{-1} h_{\bar{1}} \acts \lower0.3in\hbox{\includegraphics[scale=.4]{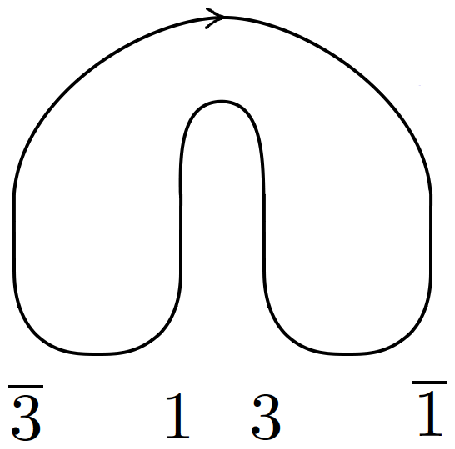}}.
\eeq
Since the braiding map intertwines the translation symmetry, this twisted product defines by construction a translation invariant function of the metric variables. 
The idea is then to deduce from  this an invariance of the {\sl non}-twisted  product of loops under a {\sl twisted} translation. 
To make this explicit at the level of the action term, let us integrate (\ref{twistedloops}) over all the variables, save one in each loop: $x_{12}^1, x_{23}^2, x_{31}^3$. We are left with:
\beq 
e^{i \Tr \left[ x_{12}^1 H_{12}\right]} \, e^{i \Tr \left[ \left(h^{-1}_{\bar{2}} h_{\bar{1}}  x^2_{23}  h^{-1}_{\bar{1}}  h_{\bar{2}}\right) H_{23}\right]} \, e^{i \Tr \left[ \left( h_{\bar{3}}^{-1} h_{\bar{1}} x^3_{31} h^{-1}_{\bar{1}} h_{\bar{3}}\right) H_{31}\right]}
\eeq
where $H_{ij} = h^{-1}_i h_j h^{-1}_{\bar{j}} h_{\bar{i}}$ denotes the holonomy along the loop $ij \bar{j} \bar{i}$. 
The invariance of this expression under the translation $x_{ij} \to x_{ij} + \epsilon$ expresses the invariance of the product 
$$e^{i \Tr \left[ x_{12}^1 H^1_{12}\right]} e^{i \Tr \left[ x_{23}^2 H^2_{23}\right]} e^{i \Tr \left[ x_{12}^1 H^3_{31}\right]}$$
under the translation $x_{ij} \to x_{ij} + \varepsilon_{ij}(\epsilon)$, where
\beq
\varepsilon_{12}(\epsilon) := \epsilon, \quad \varepsilon_{23}(\epsilon) := h^{\minus1}_{\bar{2}} h_{\bar{1}} \, \epsilon \, h^{\minus1}_{\bar{1}} h^{\minus1}_{\bar{2}}, 
\quad \varepsilon_{31}(\epsilon) := h_{\bar{3}}^{\minus 1} h_{\bar{1}} \, \epsilon \, h^{\minus1}_{\bar{1}} h_{\bar{3}}.
\eeq
In any case, the invariance leads to the identity (\ref{InvariantAction}) , which reads here
\beq
e_{H_{12}\, h^{-1}_{\overline{1}} h_{\overline{2}} \, H_{23} \, h^{-1}_{\overline{2}}h_{\overline{1}}  h^{-1}_{\overline{1}} h_{\overline{3}}\, H_{31} \, h^{-1}_{\overline{3}}h_{\overline{1}} }{(\epsilon)} = 1.
\eeq
This equality holding for all $\epsilon$, it gives a Bianchi identity of the type of (\ref{Bianchi}):
\beq
{H_{12}\, h^{-1}_{\overline{1}} h_{\overline{2}} \, H_{23} \, h^{-1}_{\overline{2}}h_{\overline{1}}  h^{-1}_{\overline{1}} h_{\overline{3}}\, H_{31} \, h^{-1}_{\overline{3}}h_{\overline{1}} }=1.
\eeq
We thus derived a Bianchi identity from the translation invariance of the vertex and propagator functions. 
In this analysis, the `twist' elements $k_e^v$ in the Bianchi identity, geometrically interpreted as parallel transport from a fixed point to the reference point of each loop, show up in commuting the variables using the $\cD\SO(3)$ braiding. 

\

More generally, we expect an analogous algorithm  for any {\sl planar} 3-bubble; namely when the link $L_v$ of the vertex has the topology of a 2-sphere. 
Starting from the 3-bubble drawn as a braided diagram, the bottom part (product of `cups') gives by construction a translation invariant function of the metric variables $x_e^j$. The algorithm will then define a sequence of topological moves corresponding to the separation of the loops, and inducing a reordering of the variables $x_e^j$, and associate to it a certain $\cD\SO(3)$ braiding map. This braiding map encodes the behavior of the amplitude (\ref{integrand}) under non-commutative translation. 
The example shown above is particularly simple, as the braided 3-bubble does not involve any crossing of the propagator strands; 
in general, an additional rule should be added in the definition of the braiding map, which would take into account such crossings.  

Just as in the above example, the action of such a braiding map on the function defined by the product of cups will induce a twisting of the variables $x_e^j$ by certain group valued functions $k_e^j(h)$ of the holonomies. A condition to obtain Bianchi identities, and hence an invariance of the action term $e^{i S_{\Delta}(x_e, h_l)}$, 
is that these functions do not depend on the variables $j$ within a loop $e$ : namely $k_e^j:= k_e$. 
We conjecture that this condition can be reached precisely when the 3-bubble is planar, namely when all the crossings of the braided diagram are removable by some topological move. 
In the presence of non-trivial crossings, on the other hand, the braiding map will give an invariance of the amplitude (\ref{integrand}) 
which will not translate into any Bianchi identity or an invariance of the action term $e^{i S_{\Delta}(x_e, h_l)}$ (obtained from (\ref{integrand}) 
by integration over all the variables $x_e^j$,  save one per loop). 
This reflects a breaking of the discrete diffeomorphism symmetry whenever the (closed) GFT graph has non-spherical 3-bubbles, 
namely for pseudo-manifold graphs. 
 
Whether this conjecture can be proven remains to be seen: we leave this for future work. 
It will also be important to understand how this analysis is affected by the use of a non-trivial braiding in the algebra of GFT fields, intertwining  the quantum symmetry.

\subsection{Open diagrams and n-point functions}



The geometrical and algebraic analysis of the previous two sections can be extended to {\sl open}  GFT graphs, with fixed boundary metric or connection data. 
An open GFT graphs is dual to a simplicial complex with boundaries. 
We have seen that the invariance of the Feynman integrand (exponential of the action) under non-commutative translation of a vertex $v$ of this simplicial complex is due to a discrete Bianchi identity  on the link $L_v$ of the vertex. 
We showed both geometrically and algebraically that, when $v$ is in the bulk, the invariance holds only when  $L_v$ has a trivial topology, or equivalently, when the 3-bubble associated to $v$ is planar. 

The same condition applies when the vertex lies at the boundary. In this case, the link $L_v$ defines an open surface, whose boundary is a loop circling the vertex $v$: 
this is the link $\partial L_v$ of $v$ in the boundary triangulation. Now, in the `group' representation, the boundary data encodes a boundary connection. 
One can then easily convince oneself that a discrete `Bianchi identity' on the link $L_v$ simply says that the holonomy of this boundary connection along $\partial L_v$ is trivial. Such a discrete Bianchi identity, and hence the invariance of the Feynman integrand under non-commutative translation of the vertex, 
hold when the link $L_v$ has a trivial (disk) topology. 

We had already noticed, at the level of the GFT vertex,  that our symmetry implies (in the group representation) flatness of the boundary connection. 
In fact, dealing with a flat boundary connection means that the holonomies along all (3d) contractible loops are trivial. 
Now, the loop $\partial L_v$ circling the boundary vertex $v$ is contractible precisely when the link $L_v$ has a trivial topology; 
the invariance under translation then holds and expresses precisely that the holonomy is trivial.   
Thus, the behavior of the Feynman integrand under non-commutative translation of the boundary vertices indeed encodes the flatness of the boundary connection, 
namely what we expect as a result of diffeomorphism invariance. 

More generally, for the GFT graphs dual to manifolds, the behavior of the Feynman amplitudes under our quantum GFT symmetry 
is consistent with what we know about discrete diffeomorphisms at the quantum level from canonical (discrete) 3d gravity as well as its covariant path integral formulation. 
Since not much is known about the action of diffeomorphisms in simplicial gravity on pseudo-manifold, we conclude that we are not missing any expected feature of discrete diffeomophism invariance, in our trivially-braided GFT formalism, as far as it can be seen at the present stage of development.

\

Given the interpretation of our GFT symmetry as the counterpart of diffeomorphism invariance, it is natural to ask whether the GFT n-point functions respect the symmetry. We know this is not the case: sticking to the usual GFT formalism, we have used a trivial braiding in the algebra of fields, 
which does not commute with the action of our symmetry transformations. As it is well-known, this leads  generically to a breakdown of the symmetry at the quantum level. In the context and spirit of the braided quantum field theory formalism, it would be more natural to use a non-trivial braiding intertwining the symmetry 
and hence fully implement the covariance of the n-point functions.  However the consequences of such a non-trivial braiding -- though currently under investigation -- are difficult to forecast, at this stage. In fact, it should clear from the above analysis of the amplitudes that the properties of GFT n-point functions in this trivially-braided GFT context do not seem to indicate inconsistencies, nor a specific physical reason why a non-trivial braiding would be necessary, or any problem with the implementation of diffeomorphism invariance. On the contrary, none of the expected features of diffeomorphisms seems to be missing in this formalism.

\section{Diffeomorphisms in topological models in higher dimensions}
\label{4dBFcase}

The analysis of the previous sections can be extended  to higher dimensions, for models describing BF theory, in a rather straightforward manner. 
Here we consider the the Ooguri GFT \cite{ooguri} for 4d BF theory, generalized to include colors. 
The variables are  complex scalar fields $\vphi_\ell$, with $\ell=1,..,5$ defined on $G^{\ot 4}=\SO(4)^{\ot 4}$, 
which satisfy the gauge invariance condition:
  \beq \label{gauge4d}
  \forall h\in \SO(4), \quad \vphi_\ell (h g_1, h g_2,h g_3,h g_4)= \vphi_\ell ( g_1,  g_2, g_3, g_4) \quad \forall \ell.
  \eeq 
The action of the model is $S[\vphi]=S_{kin}[\vphi]+S_{int}[\vphi] $ with  
\bes\label{ooguri}
S_{kin}[\vphi]&=& \int [\extd g]^4\, \sum_{\ell=1}^4\vphi_\ell(g_1,g_2,g_3, g_4) \ov\vphi_\ell(g_1,g_2,g_3, g_4)\\
S_{int}[\vphi] &=&  \lambda \int [\extd g]^{10}\, \vphi_1(g_1,g_2,g_3, g_4) \vphi_2(g_4,g_5,g_6, g_7) \vphi_3(g_7,g_3,g_8, g_9) \vphi_4(g_9,g_6,g_2, g_{10}) \vphi_5(g_{10},g_8,g_5, g_1) \nn\\
&+&  \lambda\int [\extd g]^{10}\, \ov\vphi_5(g_{1},g_5,g_8, g_{10})\ov\vphi_4(g_{10},g_2,g_6, g_9)\ov\vphi_3(g_9,g_8,g_3, g_7) \ov\vphi_2(g_7,g_6,g_5, g_4) \ov\vphi_1(g_4,g_3,g_2, g_1).    \nn
\ees
\begin{figure}
\begin{center}
\includegraphics[scale=.3]{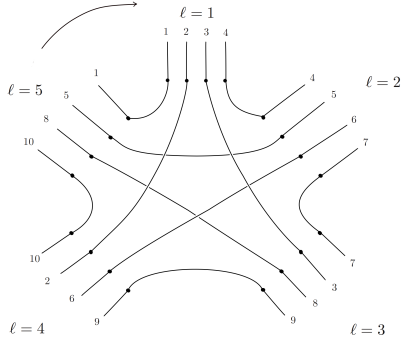} 
\label{4-vertex}
\psfrag{g}{$g_1$}
\caption{4d GFT vertex}
\end{center}
\end{figure}

Just as in 3d,  the above structures have a natural simplicial interpretation. The field $\vphi_\ell(g_1,..,g_4) $ represents a $3$-simplex (tetrahedron),  its four arguments being associated to its boundary triangles. The interaction encodes the combinatorics of five such tetrahedra glued pairwise along common triangles to form a $4$-simplex. The kinetic term  encodes the glueing of $4$-simplices along shared $3$-simplices.  

The group Fourier transform giving the metric representation is easily extended \cite{aristidedaniele} to  functions of (several copies of) $\SO(4) \sim \SU(2)\times \SU(2)/\mathbb{Z}_2$,
\beq
\hvphi_\ell(x_1,..,x_4)\equiv \int[\extd g]^4\, \vphi_\ell(g_1,..,g_4)\, \e_{g_1}(x_1)..\e_{g_4}(x_4), \quad x_i\in \so(4)\sim \R^6.
\eeq
The plane waves $\e_g \mapsto \so(4) \sim \R^6 \to \U(1)$ are defined as the product of $\SU(2)$ plane waves defined in Section \ref{introGFT},
using the decompositions $g=(g_\minus, g_\plus)$ and $x=(x_-,x_+)$ of the group and $\so(4)$-algebra elements into left and right components: 
\beq
\e_g(x) = \e^{i\Tr x_\minus g_\minus}  \e^{i\Tr x_\plus g_\plus}.
\ee
The $\star$-product is  the Fourier dual of the convolution product of $\SU(2)$ introduced in Section \ref{introGFT}. 
The variables $x$ are geometrically interpreted as  {\sl bivectors} that the standard lattice BF theory assigns to triangles, in each tetrahedron. 
Just as in 3d, the gauge invariance condition (\ref{gauge4d}) is dual, upon Fourier transform, to a closure constraint 
$\widehat C(x_1,..x_4) \!=\! \delta(\sum_{i=1}^4 \,x_i)$ of the four field variables, imposed by a non-commutative delta function  defined as in (\ref{delta}).
 


By extending the 3d symmetry analysis to the 4d case,  we will consider the action on of rotations and  translations of the quantum double\footnote{Note that a priori we could choose a bigger quantum group, like a deformation of the Poincar\'e group in six dimensions. The classification of quantum symmetries for non-commutative spaces has  been  only partially completed in 4d \cite{zakrewski}. Deformations of symmetries for higher dimensional spaces have  still to be explored. In our case, the choice of quantum group of interest is dictated  by the kinematical phase space of 4d BF theory and by its known discrete classical symmetries, which we want to encode at the GFT level.} $\cD\SO(4)$ on the scalar fields  $\vphi_\ell$. The action of the double on fields over the group is the same we presented in Section (\ref{sec:dsu2}). 
Thus an element $f \otimes \Lambda$, with $f\in \cC(\SO(4))$ and $\lambda \in \SO(4)$ acts on 
a function $\phi\in \cC(\SO(4))$ as
\beq
\phi(g)\dr \phi(\Lambda\mone  g \Lambda),  \quad \phi(g)\dr f(g)\phi(  g)
\eeq
and dually on its group Fourier transform $\widehat{\phi}(x)$ by conjugacy and translation of the Lie algebra  variable $x$. 

\medskip 

As in the Boulatov case, we easily check that the only gauge covariant action of rotations which leave the interaction term invariant is the diagonal rotation: 
In the metric formulation, gauge covariance simply means that a rotation preserves the closure $\delta(\sum_{i=1}^4 \,x_i)$ of the bivectors.

\begin{figure}[h]
\begin{center}
\includegraphics[scale=.2]{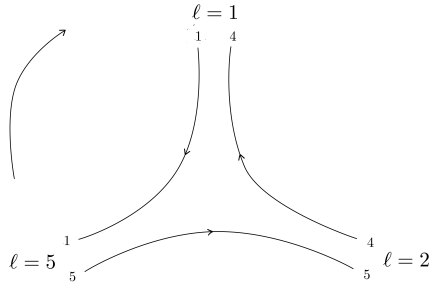} 
\caption{Vertex diagram for the edge $(34)$.  }\label{edge-diag}
\end{center}
\end{figure}

The realization of the translation symmetries is analogous to 3d, except that now they act at the {\sl edges} of the simplices patterned by the fields, rather than the vertices. The transformations are thus generated by four $\so(4)$-translation parameters $\varepsilon_e$, where $e$ labels the ten edges of the interaction 4-simplex, diagrammatically represented by its dual diagram in Fig. 9.  
Each edge of this 4-simplex is represented by an subdiagram called `edge  graph' . Thus, if $e_{\ell\ell'}$ denotes the unique edge that does {\sl not} belong to 
the tetrahedra $\ell$ or $\ell'$, the edge graph associated to $e_{\ell\ell'}$ is obtained by  removing all the lines which contain strands of color $\ell$ or $\ell'$.  
The edge graph of $e_{34}$ is pictured Fig. \ref{edge-diag}: its three lines represent the three triangles $1, 3, 4$ sharing  $e_{34}$. 

To define the action of a translation of the edge $e_{34}$, we equip the lines of the edge graph with an orientation, as drawn in the figure. Using this convention, each line has an `incoming' and an `outcoming' external strand. 
A translation of $e_{34}$ generated by  $\epsilon_{34} \in \so(4)$ acts non-trivially only the strands of the edge graph. 
In the metric representation, it  shifts the corresponding  variables $x_i^\ell$ by $\pm \epsilon_{34}$ whether the strand $i$ comes in or out of $\ell$:
\bes
&& x_i\mapsto x_i - \eps_{34} \quad \textrm{if } i \textrm{ is outgoing} \nn\\
&& x_i\mapsto x_i + \eps_{34} \quad \textrm{if } i \textrm{ is incoming}. \label{edge translation}
\ees
in a way that preserves the closure $\delta(\sum_{i=1}^4 x^\ell_i)$ of each tetrahedron.  
More precisely, the translation $\cT_{\eps_{34}}$ of the edge $e_{34}$ acts on the dual fields as 
\bes \label{VertexTranslation-edge}
&& \cT_{\eps_{34}} \acts \hvphi_1(x_1, x_2, x_3, x_4) =  \bigstar_{\eps_{34}}  \hvphi_1(x_1-\eps_{34}, x_2, x_3, x_4+\eps_{34})\nn
\\ &&
\cT_{\eps_{34}}  \acts \hvphi_2(x_4, x_5, x_6, x_7) =  \bigstar_{\eps_{34}} \hvphi_2(x_4-\eps_{34}, x_5+\eps_{34}, x_6, x_7)\nn
\\&&
\cT_{\eps_{34}}  \acts \hvphi_5(x_{10}, x_8, x_5, x_1) =   \bigstar_{\eps_{34}}  \hvphi_5(x_{10}, x_8, x_5-\eps_{34}, x_1+\eps_{34})\nn
\\&&
\cT_{\eps_{34}} \acts \hvphi_\ell = \hvphi_\ell  \qquad \textrm{ if } \ell = 3,4.
\ees
The same field transformation is expressed in a more explicit way (without star product) in the group representation, as follows: 
\bes \label{translation def}
&& \cT_{\eps_{34}} \acts \vphi_1(g_1, g_2, g_3, g_4) = \e_{g_1^\inv g_4}(\eps_{34})  \vphi_1(g_1, g_2, g_3, g_4)\nn
\\ &&
\cT_{\eps_{34}}  \acts \vphi_2(g_4, g_5, g_6, g_7) = \e_{g_4^\inv g_5}(\eps_{34})  \vphi_2(g_4, g_5, g_6, g_7)\nn
\\&&
\cT_{\eps_{34}}  \acts \vphi_5(g_{10}, g_8, g_5, g_1) = \e_{g_5^\inv g_1}(\eps_{34})  \vphi_5(g_{10}, g_8, g_5, g_1)\nn
\\&&
\cT_{\eps_{34}} \acts \vphi_\ell = \vphi_\ell \qquad \textrm{ if } \ell = 3,4.
\ees


We see that this transformation matches the intuition corresponding to translating bivectors ($\mathfrak{so}(4)$ Lie algebra elements) associated to the triangles 
of the 4-simplex dual the GFT interaction vertex, by means of Lie algebra valued generators associated to its edges. 
This matches also the action of diffeomorphisms on the bivectors of discrete BF theory (recall that the transformations we have defined take the closure condition (metric compatibility) into account)\footnote{Unlike the 3d case, however,  we have no geometric description in terms of translating edges of a 4-simplex embedded in four-dimensional flat space. This is only to be expected, given that we are dealing with a non-geometric theory, and thus with non-geometric 4-simplices.}. 
It can be checked by direct calculation that the GFT action \eqref{ooguri} is invariant under the above field transformations. 

In fact, one verifies, as in the 3d case, that both kinetic and vertex {\sl  functions} themselves are left invariant -- before integration. 
A way to make this invariance manifest is to extract from, say, the vertex function in group variables, the conservation laws for the holonomies associated to edges of the 4-simplex dual to the GFT vertex. Just as in Sec. \ref{sec:vertex-variables}, the explicit integration over the group elements $h_\ell$ in the vertex gives: 
\beq \label{interac-holonomies}
V(g,g') = \int \prod_{i=1}^5 [\extd h_\ell]\, \prod_{i=1}^{10} \delta(g^\ell_ih_\ell h_{\ell'}\mone ({g^{\ell'}_i})\mone) 
 = \delta(G_{12}) \delta(G_{13}) \delta(G_{15})  \delta(G_{23}) \delta(G_{25}) \delta(G_{35}),
 \eeq 
where:
\bes 
&&
G_{12} =  g_8 {g_9}\mone g'_9 g_{10}\mone g'_{10} {g'_8}\mone\qquad  
G_{13}=  g_5   g_6 \mone{g'_6} g_{10}\mone g'_{10} {g'_5}\mone, \qquad  
G_{15}= g_7g_6\mone {g'_6} {g'_9}\mone g_9 {g'_7}\mone, \nn\\
&&
G_{23}= g_2\mone g'_2g_{10}\mone g'_{10} {g'_1}\mone g_1, \qquad  
G_{25}=  {g'_3}\mone g_3g_{2}\mone g'_{2} {g'_9}\mone g_9, \qquad  
G_{35}= g_4g_2\mone g'_2 {g'_6}\mone g_6 {g'_4}\mone.
\nn
\ees
We recognize here the $G_{ij}$ as the holonomies around the edges $(ij)$. The delta functions in \eqref{interac-holonomies} encode the flatness conditions which, as expected from the canonical analysis of discrete BF theory, constrain the connection variables as a result of the diffeomorphism symmetry.

Note that the holonomies associated to the edges $(i4)$ are missing. This is analogous to  the 3d case where the translations of the 4 vertices of the tetraedron are not all independent, only 3 of them are. Also, in the 4d case, the translations of the edges are not all independent, just as the continuum symmetry can be shown to be reducible (\cf Appendix \ref{app}): this is due to the Bianchi identities satisfied by the boundary connection represented by the field variables. 
In fact one can prove that translating a vertex, \ie translating  all  edges sharing this vertex,  leaves invariant the interaction term, and by extension the integrand of the Feynman amplitude. 
The true symmetry is therefore represented by the above edge translations {\it modulo} the translations of the edges following a vertex translation. 

\medskip

We thus see that, for (the GFT model describing) 4d BF theory, everything proceeds in parallel with the 3d case,  the only new ingredient being the reducibility of the resulting symmetry. However the strategy used here to define the action of diffeomorphisms in GFT can in principle be extended to the  physically more interesting case of 4d gravity GFT models, obtained by constraining the topological one \cite{review, gftreview}. 
In general, we  expect that the imposition of the simplicity constraints will break the full symmetries of Ooguri's GFT.  
It will be interesting to determine whether there is an eventual remnant symmetry, and if not, whether the vertex translations become then the relevant, if only approximate, symmetry \cite{4ddiffeo}. In this case, such a symmetry could admit a good geometric interpretation as  translations of the vertices of a geometric 4-simplex in an embedding 4d flat space, as we expect from diffeomorphisms in discrete gravity \cite{bianca, bennybianca}.

\section{Additional insights}
\label{additional}

We now discuss additional insights that the newly identified GFT symmetry provides, concerning various aspects of the GFT formalism itself. While these are somewhat secondary results, we believe they confirm the importance of the new symmetry and suggest that further progress can be triggered by its identification.

\

{\bf The necessity of coloring.}\\
The introduction of {\it coloring} in GFT models in \cite{razvan} has already been proven useful in studies of the topological properties of the Feynman diagrams generated by such models \cite{vincentcolored, razvan2, razvan3}, in particular for the automatic removal of complexes with some types of extended singularities that are instead generated by the non-colored models. Most important, it has been crucial for the proof that the 3d GFT we have studied admits a topological expansion of its Feynman diagrams such that manifolds configurations of trivial topology dominate the sum for large values of the representation cut-off \cite{RazvanN}. These important results have important implications for the program of GFT renormalization, and for defining a GFT generalization of the notion of (double) scaling limit of matrix models \cite{FGO}, and thus for the understanding of the continuum limit. No obvious physical or geometric relevance, however, had been discovered, until now, for the same coloring. Our results show, on the other hand, that coloring is a {\it necessary} feature of GFT models for 3d gravity and general BF theories. In fact coloring is  a necessary ingredient in the definition of the GFT diffeomorphism symmetry we have identified and discussed in this paper. More precisely, it can be shown that removing the coloring leads to the immediate breaking of the symmetry and that only a restricted translation of the vertices of the tetrahedron dual to the GFT vertex can be defined as a field transformation leaving the non-colored action invariant, such as the one identified in \cite{Floetera}. This symmetry however, although being a particular combination of the symmetry transformation we have studied, does not have a clear simplicial gravity interpretation. Given the interpretation of our GFT symmetry as the counterpart of discrete diffeomorphisms in simplicial gravity path integrals, the importance of coloring from the physical/geometrical point of view becomes instead manifest. In its light, we recognize the colored Boulatov GFT model as {\it the} correct GFT description of 3d quantum gravity.

\

{\bf A braided group field theory formalism?}\\
In this paper, we have studied the issue of diffeomorphism symmetry within the standard (colored) group field theory formalism. In particular, the algebra of fields we have worked with has been assumed to have {\it trivial braiding} \cite{majid, oeckl}, i.e. the map between the tensor product of two fields and the one with opposite ordering is given by the trivial flip map. At the same time, however, the symmetry we have identified in the GFT action corresponds, as we have stressed, to a {\it quantum group} symmetry acting on this space of fields. As such its action on the space of fields would naturally induce, when these are defined as elements in its representation category, a non-trivial braiding structure \cite{SasaiSasakura}. This also results in a corresponding braided statistics \cite{oecklStat}. Most important, it can be shown that the use of the induced braiding map in the algebra of fields is necessary, if the symmetry is to be preserved at the quantum level \cite{SasaiSasakura, oeckl2, mangano}, for example so that the correct Ward identities for n-point functions follow from the existence of the symmetry at the level of the action. We will discuss briefly below whether this is necessary on physical grounds in our context, and what the properties of the n-point functions are  in our trivially-braided context. In any case, the above considerations suggest, at least from a mathematical and field theoretic perspective,  to consider a generalization of the GFT formalism, beyond the one as non-commutatve field theories, achieved in \cite{aristidedaniele}, to a {\it braided non-commutative group field theories} (see also \cite{Floetera} for further arguments in this direction.). The first issues to tackle, in this direction, are: 1) what is the correct braiding among GFT fields intertwining our quantum group symmetry, if it exists at all; 2) what are the physical consequences of the implementation of a non-trivial braiding and of the resulting quantum Ward identities, from the point of view of simplicial quantum gravity, loop quantum gravity and spin foam models. 

\

{\bf Constraints on GFT model building}\\
Another useful role that symmetries play in usual quantum field theories is that they help constraining the allowed field interactions. In fact, the requirement that the GFT interactions preserve the quantum group symmetry we identified as discrete diffeomorphisms rules out some GFT interactions that could be considered, a priori,  as admissible.

We have already discussed above how removing the coloring form the GFT fields, i.e. considering the original Boulatov formulation with a single field, breaks the symmetry. This can also be understood as a special case of a larger set of possible GFT interactions {\it within the colored GFT formulation}, that we now see to be ruled out by symmetry considerations. The colored model we worked with, for 3d gravity, was based on 4 different fields $\varphi_l$, with $l=1,2,3,4$, and the only interaction term was of the form $\varphi_1\,\varphi_3\,\varphi_2\,\varphi_4$ (plus complex conjugate), with standard tetrahedral combinatorics of arguments. The single-color Boulatov interaction corresponds to terms of the type $\varphi_l\,\varphi_l\,\varphi_l\,\varphi_l$. A quick calculation shows that not only such terms, but any interaction being more than linear in any of the colored fields (e.g. $\varphi_1\,\varphi_1\,\varphi_3\,\varphi_4$ or  $\varphi_1\,\varphi_2\,\varphi_3\,\varphi_2$)  is not invariant under our GFT diffeomorphism symmetry. 
We are then left with interactions that involve linearly all the d GFT fields (in models generating d-dimensional simplicial structures). The ordering of such fields can be chosen at will (in our trivially-braided context). 

We can however also ask whether the {\it ordering of the (group or non-commutative) arguments} of such fields in the interaction term can be chosen at will. Different orderings, in fact, have been considered in the literature (see \cite{FGO}). In colored models the order of the arguments of the field is considered as fixed and does not play any special role (the orientability of the resulting Feynman diagrams is already ensured by the complex structure and by the requirements of same-color propagation only \cite{razvan3,francesco}). Regarding the interplay between ordering of arguments and symmetry, the situation is slightly trickier. It can be seen easily that, for any given choice of ordering, there exists a (set of) transformation(s) acting on the $d+1$ fields leaving the action invariant and corresponding to diffeomorphisms, in the sense we have discussed. The very definition of the transformations retains the imprint of the chosen ordering of field arguments.  At the same time, however, it can be shown that such transformation would not, in general, leave invariant a vertex defined by a different ordering nor an action involving a sum over different orderings. This means, for example, that the GFT field itself cannot be defined to be invariant under permutations of its arguments, as this imposition would break its covariance under the diffeomorphism transformation, and then the invariance of the action. It must be said, however, that a possible way out of this restriction could be, once more, an appropriate {\it braiding} that relates fields defined with different orderings of their arguments, and possibly, intertwines our symmetry. We leave this for future work.

Last, one could consider defining both higher order interaction terms, i.e. terms of order higher than $d+1$ involving colored fields, with various choices of pairing of field arguments, as well as other terms still of order $d+1$, but defined by non-tetrahedral combinatorics of arguments. Our symmetry constrains severely model building of this type. We have not performed yet a complete analysis. However, we have considered some examples. One interesting example of alternative interaction term, the so-called \lq\lq pillow\rq\rq term has been introduced in \cite{laurentdavidTop} and studied further in \cite{vincentmatteo}. It has the following form (in its colored version):

\begin{equation}
+\frac{\lambda\,\delta}{4!}\prod_{i=1}^{6}\int dg_i\left[
\, \phi_1(g_1,g_2,g_3)\phi_2(g_3,g_4,g_5)\phi_3(g_4,g_2,g_6)\phi_4(g_6,g_5,g_1)\right].
\end{equation}

So it is given by the same type of vertex function, i.e. a product of delta functions on the group, enforcing the identification of edge variables among four triangles, as in the standard tetrahedral term. However, the combinatorial pattern is now different, and corresponds to two pairs of triangles glued to one another along two edges in each pair, and along one single edge between the two pairs. The interest in the addition of such term lies in the fact that it turns the (non-colored) Boulatov model into a Borel summable one (with some restrictions on the coupling constant $\delta$), and with a different (worse) scaling behaviour. It can be proven, however, that this term is not invariant under GFT diffeos, and thus is not an admissible modification of the action of the model, in the colored case.

We stress again that the above considerations would be modified by the introduction of a non-trivial braiding among fields, with a corresponding generalization of the GFT formalism. However, not knowing the correct braiding structure, it is impossible to be more definite about what the modifications would be.

\section*{Conclusions}
Using the recently introduced non-commutative metric formulation of group field theories, we have identified a set of GFT field transformations, forming a global quantum group symmetry of the GFT action, and corresponding to translations of the vertices of the simplices dual to the GFT interaction vertex, in a flat space embedding. The analysis of the action of these transformations at the level of the GFT Feynman amplitudes, which are given, in this metric formulation, by simplicial gravity path integrals, shows that the transformations we identified correspond to (the discrete analogue of) diffeomorphisms for fixed simplicial complex satisfying manifold conditions, and leave the same amplitudes invariant thanks to discrete Bianchi identities, whose GFT origin we are now able to exhibit. Moreover, for open Feynman diagrams dual to simplicial manifolds with boundaries, we have shown that the same transformations enforce the flatness of the boundary connection, and thus encode the simplicial version of the canonical gravity constraints, as expected. 

While we focus on the case of 3d riemannian gravity, we also show how our results generalize straightforwardly to BF theories in higher dimensions. Thus our results on the one hand match those obtained, concerning discrete diffeomoprhisms, in the context of simplicial gravity (e.g. Regge calculus), on the other hand improve them by both embedding them within a more general context and re-phrasing them in purely (quantum) field theoretic language. 
An immediate advantage of this embedding is the clear way in which we can now link to one another various aspects of diffeomorphism invariance in spin foam models, canonical loop quantum gravity and simplicial  gravity, previously discussed in the literature, and now understood to be all consequences and manifestations of the same GFT field symmetry:  the symmetry of the Regge action and the simplicial Bianchi identities (manifest in the metric representation of GFTs), the canonical constraints of loop quantum gravity (adapted to a simplicial complex) (best seen in the group picture) and the algebraic identities satisfied by $nj$-symbols and at the root of topological invariance of state sum (spin foam) models (obtained from the GFT symmetry in representation space).

Our analysis also provides some new insights on the GFT formalism itself. These include: the need for {\it coloring} in the GFT formalism, from the point of view of simplicial gravity symmetries; the possible role of {\it braiding} in this class of models, and thus in simplicial gravity path integrals and spin foam models, and the potential interest in a {\it braided group field theory} formalism; the issue of Ward identities and the relation of the same with canonical quantum gravity constraints and recursion relations for 6j- and 10j- symbols; the use of the GFT symmetries we identified for constraining the possible interaction terms that can be added to the standard GFT vertex.

We believe that the GFT symmetry we identify can also play a useful role concerning ongoing work on GFT renormalization and, possibly, for the extraction of continuum gravity from GFT models.

\section*{Acknowledgements}
We thank V. Bonzom, S. Carrozza, B. Dittrich, E. Livine, R. Oeckl and A. Perez for useful comments and discussions.
DO gratefully acknowledges financial support from the A. von Humboldt Stiftung through a Sofja Kovalevskaja Prize.

\appendix

\section{BF action and its symmetries}\label{app}
In this appendix, we recall the standard basic facts about the symmetries associated to the $BF$ action.

We work with a $d$ dimensional manifold $\mm$, equipped with a principal bundle assiciated with the semi-simple Lie group $G$. The Lie algebra of $G$ is noted $\g$ and is equipped with a non-degenerate Killing form which we note $\tr$.  $A$ is the connection, \ie a $1$-form with value in $\g$,   of the principal bundle and we note $F= dA+A\wedge A$  the curvature $2$-form of the connection $A$. $d_A$ is the covariant derivative defined in terms of the connection $A$. We now introduce $B$  a $(d-2)$-form with value in a Lie algebra $\g$.   The $BF$ action is built using the Killing form $\tr$ on $\g$.
\beq
S_{BF}=\int \tr (B\wedge F).
\eeq
The equations of motion are 
\beq\label{eom}
d_AB=0\qquad F=0. 
\eeq
The action is invariant under both translation of the $B$-field and the gauge transformations.  
The infinitesimal gauge transformations are given by
\bes\label{L}
&& A\dr A+ \delta^L_X A = A+ d_AX  = A + dX+ [A,X],\nn\\ 
&&  B\dr  B+ \delta^L_X B=B+  [B,X], \qquad 
\ees
$X\in\g$ is a scalar field with value in $\g$.
The $B$ field  is therefore transforming under the adjoint action of $G$. The curvature $F$ is also transforming under the adjoint action and it is thus easy to check that   the action is invariant under these gauge transformations.  

Thanks to the Bianchi identity $d_AF=0$, the action is also left invariant  if we translate the $B$ field by  $d_A\Phi$ where $\Phi$ is a $(d-3)$-form with value in $\g$. 
\bes\label{T}
&&  A\dr A+ \delta^T_\Phi A = A,\nn\\
&&  B\dr  B+ \delta^T_\Phi B=B+ d_A\Phi = B + d\Phi+ A\wedge \Phi. 
\ees
There is however a possible redundancy for the translations if $d\geq 4$. Indeed, assuming $d\geq4$, consider the $d-4$-form $V$ with value in $\g$, then  $\Phi$ and $\Phi'= \Phi+ d_A V$ generates \textit{on shell} the same transformation since $d_A\Phi= d_A\Phi'$, due to $d_A^2V=[F,V]$. This last term is zero on shell.  
\medskip

The $BF$ action is clearly invariant under the diffeomorphisms since it is purely topological.  Let us consider explicitely the (infinitesimal) action of the diffeomorphisms. Considering a vector field $\xi$, the infinitesimal action of the diffeomorphisms is given by the Lie derivative $\cL_\xi$. We have therefore 
\beq
B\dr \cL_\xi B= d(\iota_\xi B) + \iota_\xi (d B), \quad A\dr \cL_\xi A= d(\iota_\xi A) + \iota_\xi (d A),
\eeq
 where we have introduced the interior product $\iota_\xi$  which satisfies in particular 
 \beq\label{interior product}
 \iota_\xi (\omega_1\wedge \omega_2)= \iota_\xi (\omega_1)\wedge \omega_2+ (-1)^p\omega_1\wedge \iota_\xi (\omega_2),
 \eeq
 with $\omega_1$ and $\omega_2$ respectively  a $p$- and a $q$-form. These transformation can actually be related to the previous  transformations \eqref{L} and \eqref{T}.  
We have that 
\bes
&&\iota_\xi(d_A B)=  \iota_\xi(dB) +\iota_\xi(A\wedge B)= \iota_\xi(dB) +[\iota_\xi(A),B] -A \wedge \iota_\xi (B)\nn\\
&&\iota_\xi( F)=  \iota_\xi(dA) +\iota_\xi(A\wedge A)= \iota_\xi(dA) + [\iota_\xi(A),A]. 
 \ees
Taking $X=\iota_{\xi} B$ and $\Phi=\iota_\xi A$,  
we can reexpress the action of the diffeomorphisms as  
 \bes
&& \cL_\xi B= \delta^L _{\iota_{\xi} A} B + \delta^T _{\iota_{\xi} B} B+ \iota_\xi (d_AB)\nn\\ 
&&  \cL_\xi A= \delta^L _{\iota_{\xi} A} A + \delta^T _{\iota_{\xi} B} A+ \iota_\xi (F).
 \ees
 This means that on-shell \eqref{eom}, the diffeomorphisms action is equivalent to the translation \eqref{T} and gauge transformation \eqref{L}. 
 

\end{document}